\documentclass[journal]{IEEEtran}

\usepackage{cite}
\usepackage[table]{xcolor}
\usepackage{color}
\usepackage{subcaption}
\usepackage{ amssymb }
\usepackage{epsfig}
\usepackage{epstopdf}
\usepackage{amsmath}
\usepackage{color}
\usepackage{float}
\usepackage{enumerate}
\usepackage{placeins}
\usepackage{algorithm}
\usepackage{algpseudocode}
\usepackage{xpatch}
\usepackage{amssymb}
\usepackage{eurosym}
\usepackage{amstext} 
\DeclareRobustCommand{\officialeuro}{%
  \ifmmode\expandafter\text\fi
  {\fontencoding{U}\fontfamily{eurosym}\selectfont e}}
\DeclareCaptionLabelSeparator{periodspace}{.\quad}
\DeclareMathOperator*{\argmax}{arg\,max}

\usepackage{mathtools}

\DeclarePairedDelimiter\floor{\lfloor}{\rfloor}
\newcommand{\mb}[1]{multi-user beamforming}
\newcommand{\Mb}[1]{multi-user beamforming}
\newcommand{\be}[1]{\begin{equation}}
\newcommand{\ee}[1]{\end{equation}}
\newcommand{\beq}[1]{\begin{eqnarray}}
\newcommand{\eeq}[1]{\end{eqnarray}}
\newcommand{\bem}[1]{\begin{multline}}
\newcommand{\eem}[1]{\end{multline}}

\begin{document}
\title{Total Cost of Ownership Optimization for Direct Air-to-Ground System Design}

\author{Ergin~Dinc, \thanks{This paper was accepted by IEEE Transactions on Vehicular Technology. DOI: 10.1109/TVT.2021.3103634. Copyright \textcopyright 2015 IEEE. Personal use of this material is permitted. However, permission to use this material for any other purposes must be obtained from the IEEE by sending a request to pubs-permissions@ieee.org.}\thanks{A preliminary version of this work \cite{dinc17} was presented in IEEE GLOBECOM'17.}\thanks{E. Dinc is currently with the Cavendish Laboratory, University of Cambridge, CB3 0HE,UK. During the time of this study, he was a postdoc at KTH Royal Institute of Technology, Stockholm 164 40, Sweden. (e-mail: ed502@cam.ac.uk).}\thanks{M. Vondra is currently with Skoda Auto a.s., GG2 Department, Mladá Boleslav, Czech Republic. During the time of this study, he was a postdoc at KTH Royal Institute of Technology, Stockholm 164 40, Sweden. (e-mail: michal.vondra2@skoda-auto.cz).}\thanks{C. Cavdar is with the Center for Wireless Systems (wireless@kth), Department of Communication Systems, School of Information and Communication Technology, KTH Royal Institute of Technology, Stockholm 164 40, Sweden (e-mail: cavdar@kth.se).} Michal~Vondra, Cicek~Cavdar\thanks{This study is funded by EIT Digital ICARO-EU (Seamless Direct Air-to-Ground Communication in Europe) Project.} }

\markboth{}{Draft}

\maketitle
\vspace{-1cm}
\begin{abstract}
Aircraft cabin is one of the last venues without mobile broadband. Considering future 5G applications and connectivity requirements, direct air-to-ground communications (DA2GC) is the only technique which can provide high capacity and low latency backhaul link for aircraft via a direct communication link. To this end, we propose an analytical framework to investigate the ground station deployment problem for DA2GC network employing multi-user beamforming with dual-polarized hybrid DA2GC antenna arrays. In addition, the proposed framework is utilized to analyze and optimize the total cost of ownership (TCO) of the DA2GC network to provide coverage for European airspace. We present the interplay between different network parameters: the number of ground stations, array size, transmit power and bandwidth, and TCO optimizing deployment parameters are calculated in order to satisfy capacity requirements. At the end, we show that, depending on the cost of different network resources, a terrestrial cellular network can be designed to cover the whole European airspace with limited number of ground stations with a certain array size, i.e., 900 and 361 antenna elements for ground station and air station, respectively.
\end{abstract}

\IEEEpeerreviewmaketitle

\section{Introduction}

Users demand high speed broadband connectivity ubiquitously regardless of location and time. Today, the tasks requiring broadband connectivity need to be interrupted during the flight since aircraft are still left as one of the few venues without high speed internet. Although all the components of the direct air-to-ground communications (DA2GC) are available, mobile operators are hesitant to deploy such a large-scale system to cover the sky over a continent for a specific use case due to cost concerns. The cost analysis of such a system is challenging. There is an interplay between number of antennas, number of ground base stations (GSs) and bandwidth. In order to understand this interplay we need to answer the following questions: What is the ground cellular network capacity needed to cover the flight paths? What type of cellular network deployment is more cost efficient? Is it better to deploy large GSs with massive number of antennas or to limit the number of antennas by accepting the additional cost of extra GSs? What is the impact of spectrum availability and cost in these decisions? In order to answer such questions, capital expenditures (CAPEX) and operational expenditures (OPEX) need to be taken into account to have an estimate of the cost over a time span. Total Cost of Ownership (TCO) is a metric to estimate the direct and indirect costs of a certain technology or system accounting for the CAPEX and OPEX to gauge the viability of an investment. 


Recently, in-flight connectivity is getting significant attention from both industry and academia. In \cite{vtc}, the authors introduce the seamless gate-to-gate connectivity concept to provide connectivity services in all phases of flights. According to this concept, in-cabin network is powered by Wi-Fi, licensed 3G/LTE, and unlicensed LTE standards to cover all flight phases. In addition to the technical studies, business modeling for in-flight connectivity imposes new challenges since several business entities: airline, in-cabin operator, mobile network operator, terrestrial operator, satellite operator, and content operator, are required to work together to create the value for the passengers. To this end, ecosystem type business models are proposed and investigated in \cite{dinc2017commag}.

In-flight broadband (IFB) connectivity requires backhaul capacity for aircraft which can be provided via DA2GC and/or satellite communication (SATCOM). SATCOM-based solutions are natural choice when transcontinental flights are considered. However, continental flights have a significant share in the airline market. More than $800$ million passengers traveled within Europe in 2015 \cite{atag}. SATCOM based IFB solutions generally utilizes Ku-band satellites, and provide $70-100$ Mbps/aircraft \cite{gogo2ku}. However, SATCOM imposes very high transmission delays ($\approx500$ ms). In \cite{vondra17c}, the authors investigate and compare achievable data rates for different technologies: satellite communication (SATCOM) and DA2GC. As also concluded in \cite{vondra17c}, SATCOM based IFB services cannot compete with future DA2GC systems with low transmission delays ($10$ ms) and high backhaul capacities ($1.2$ Gbps) as suggested by Next Generation Mobile Networks (NGMN) Alliance \cite{ngmn}. Since {it is not possible to provide transcontinental coverage via DA2GC network}, a full-scale IFB connectivity solution, {i.e.} global coverage, requires a hybrid network including both DA2GC and SATCOM. {However, in this paper, we consider continental coverage for the European Airspace; thus, this paper only focuses on DA2GC for IFB connectivity.}

\begin{figure}[!ht]
\centering
  \includegraphics[width=0.45\textwidth]{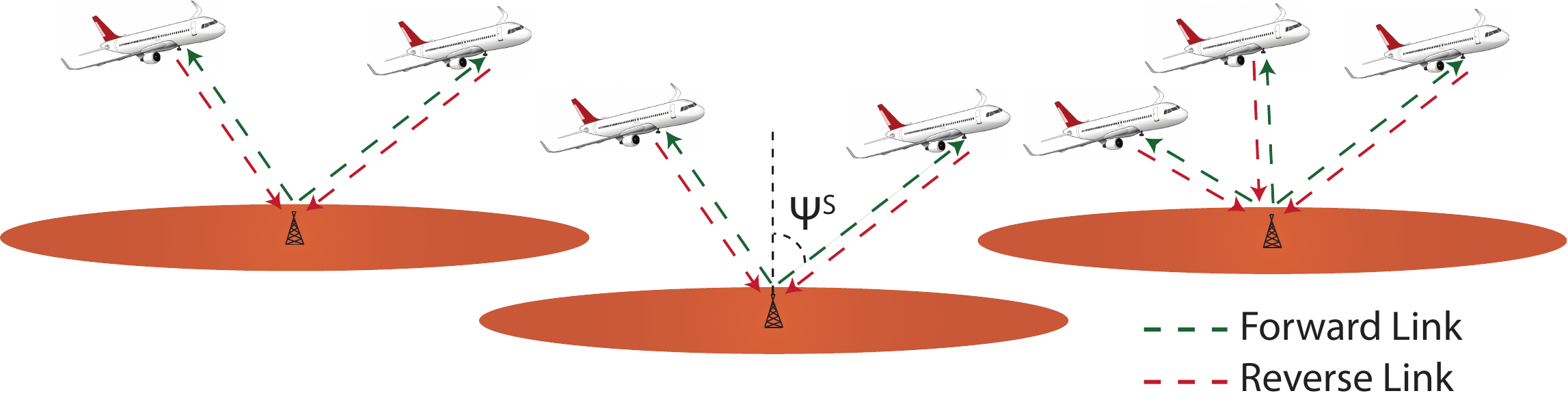}
  \caption{DA2GC. $\Psi_s$ is the transmission angle from the vertical axis of the cell center to the cell edge aircraft.}
  \label{fig:sys}
\end{figure}

In DA2GC, special ground base stations are deployed to provide connectivity to aircraft. As seen in Fig. \ref{fig:sys}, downlink (or backhaul link) for aircraft is named as forward link (ground-to-air). There are some early applications providing IFB services. More than $200$ DA2GC GS are deployed across US and Canada based on CDMA2000 technology by Gogo Inc. \cite{atg4}. However, this system is bandwidth limited (2x2 MHz) and able to provide $9.8$ Mbps/aircraft. In addition, Deutsche Telekom and Inmarsat are also deploying a DA2GC/SATCOM hybrid IFB system in S band frequencies ($2$x$15$ MHz) to provide air-to-ground (A2G) connectivity up to 75 Mbps/cell \cite{inmarsat}. Available DA2GC based IFB solutions are not able to provide high sustaining bit rates. To tackle this problem, there is an ongoing activity in Europe to allocate a designated spectrum for DA2GC \cite{ecc}. In addition, advanced communication techniques such as large antenna arrays, multi-user beam-forming and higher order modulation schemes will improve DA2GC systems to provide capacity levels estimated by the industry: $1.2$ Gbps per aircraft \cite{ngmn}.

The literature on DA2GC mostly focus on unmanned aerial vehicle (UAV) communication \cite{mozaffari17,khuwaja18,khawaja18,qureshi18}. However, the channel modeling efforts for UAVs are not directly applicable to aircraft case as characteristics of communication channel for UAVs are completely different than aircraft case. Whereas aircraft follow designated flight paths, UAVs can make fast and random manoeuvres, which cause beam-tracking problems. Therefore, beamtracking is a nearly deterministic process in the case of aircraft and ground base stations can be placed in order to ensure line-of-sight for higher reliability. There also exist some works on DA2GC for aircraft. In \cite{tadayon16}, the authors investigate the possibility of extending the existing LTE network for aircraft by using up-tilted ground stations (GSs). However, \cite{vondra17c} has revealed that the capacity levels demanded by the industry can be only met by DA2GC network. Furthermore, \cite{tang18} provides the investigation of uniform circular arrays for DA2GC links near airports, but uniform circular arrays cannot provide wide angle coverage, which is necessitated by DA2GC GSs having 50-150km range as discussed in Section \ref{secsim}. 

{DA2GC has been utilized in air traffic control (ATC) systems more than 50 years. These very high frequency (VHF) systems are mainly built for supporting voice communication with a ground terminal. In addition, the current state-of-the-art technology VHF data link (VDL2) enables A2G data transfer with $31.5$ kbps \cite{helfrick07}. Towards the future ATC solutions, European SESAR \cite{sesar}, SANDRA \cite{plass12} and American NextGen \cite{schnell04} initiatives have proposed a data centric approach based on all-IP solution and backward compatible multiple communication techniques. As a result of these initiatives, seamless ATC networking to monitor airplanes can be provided through L-band, Ku-band satellites for global coverage, Aeronautical Mobile Airport Communications System (AeroMACS) for situational awareness near airports \cite{ehammer11}, and VDL2 and VHF as back-up \cite{fernandez15}. These ATC systems are for mission critical communication and designed to achieve low data rate links. Thus, they are not suitable for IFB connectivity services.}
\begin{table*}[h!]
\caption{(A) NGMN's DA2GC KPI Requirements \cite{ngmn}, (B) Flight Statistics in Europe \cite{atag}.}
\centering
\label{reqtab}
\scalebox{0.8}{
\centering
\begin{tabular}{ccc|c|c|c|c|}
\cline{1-2} \cline{4-7}
\multicolumn{2}{|c|}{\textbf{Table I-A}}                                                                                                                                               & \multicolumn{1}{l|}{\textbf{}} & \multicolumn{4}{c|}{\textbf{Table I-B}}                                                                                                                         \\ \cline{1-2} \cline{4-7} 
\multicolumn{1}{|c|}{\textit{\textbf{Parameter}}} & \multicolumn{1}{c|}{\textit{\textbf{Requirements}}}                                                                                & \textit{\textbf{}}             & \textit{\textbf{Parameter}}                                                  & \textit{\textbf{2014 {[}3{]}}} & \textit{\textbf{2020}} & \textit{\textbf{2030}} \\ \cline{1-2} \cline{4-7} 
\multicolumn{1}{|c|}{\textit{Data rate}}          & \multicolumn{1}{c|}{\textit{\begin{tabular}[c]{@{}c@{}}Download: 15 Mbps/active user\\ Upload: 7.5 Mbps/active user\end{tabular}}} & \textit{}                      & \textit{Passengers}                                                          & \textit{873,400,000}           & \textit{1,079,870,606} & \textit{1,538,045,821} \\ \cline{1-2} \cline{4-7} 
\multicolumn{1}{|c|}{\textit{Latency}}            & \multicolumn{1}{c|}{\textit{10 ms}}                                                                                                & \textit{}                      & \textit{Flights}                                                             & \textit{7,560,360}             & \textit{9,347,619}     & \textit{13,313,693}    \\ \cline{1-2} \cline{4-7} 
\multicolumn{1}{|c|}{\textit{Aircraft Density}}   & \multicolumn{1}{c|}{\textit{60/18000 km$^2$}}                                                                                      & \textit{}                      & \textit{Total Flight Hours}                                                  & \textit{15,120,720}            & \textit{18,695,238}    & \textit{26,627,387}    \\ \cline{1-2} \cline{4-7} 
\multicolumn{1}{|c|}{\textit{Traffic Density}}    & \multicolumn{1}{c|}{\textit{\begin{tabular}[c]{@{}c@{}}Download: 1.2 Gbps/aircraft\\ Upload: 600 Mbps/aircraft\end{tabular}}}      & \textit{}                      & \textit{Aircraft in service}                                                 & \textit{6,586}                 & \textit{8,142}         & \textit{11,597}        \\ \cline{1-2} \cline{4-7} 
\textit{}                                         & \textit{}                                                                                                                          & \textit{}                      & \textit{Load factor}                                                         & \multicolumn{3}{c|}{\textit{81}}                                                 \\ \cline{4-7} 
\textit{}                                         & \textit{}                                                                                                                          & \textit{}                      & \textit{\begin{tabular}[c]{@{}c@{}}Average \# of \\ Passengers\end{tabular}} & \multicolumn{3}{c|}{\textit{200}}                                                \\ \cline{4-7} 
\textit{}                                         & \textit{}                                                                                                                          & \textit{}                      & \textit{Average flight time}                                                 & \multicolumn{3}{c|}{\textit{2}}                                                  \\ \cline{4-7} 
\end{tabular}}
\end{table*}

DA2GC network requires advanced communication techniques to satisfy the demand from passengers and industry. However, the cost analysis of the network is critical to determine the key network parameters, i.e., the number of GSs, the number of antenna elements, bandwidth and transmit power. To tackle this problem, this paper has two major contributions: (1) an analytical framework to calculate average backhaul capacity for an aircraft as a function of different network design parameters, and (2) TCO optimization for DA2GC network to calculate the network parameters that can satisfy the backhaul DA2GC link capacity requirements. Therefore, this paper provides a broad investigation of DA2GC from physical layer modeling to TCO of DA2GC network. For the first objective, we propose a novel multi-user beamforming algorithm for dual-polarized planar antenna arrays in our preliminary work \cite{dinc17}. In the beamforming algorithm, we also include the effects of beamsteering-loss, which is often neglected in the beamforming literature. Since DA2GC GSs are required to steer their beams to very low elevation angles to provide cell-edge coverage, achievable array gain significantly decreases with increasing steering angle. To tackle this problem, we introduce the utilization of multifaceted antenna arrays to minimize beamsteering-loss in DA2GC GSs for the first time in the literature. multifaceted antenna designs have been successfully utilized in tri-sectored cellular stations, millimeter wave antenna arrays \cite{pi16} and radar arrays to reduce beamsteering-loss. In this paper, we determine the optimum multifaceted antenna structure to minimize the beamsteering-loss in DA2GC GSs. {We also extend our previous work \cite{dinc17} to include beam-alignment loss, which is the amount of power-loss due to misalignment between beams from GS and aircraft. Furthermore, we investigate the effect of Doppler shift in DA2GC channels and intercell interference in the proposed DA2GC network.} At the end, we derive an analytical formula to calculate average achievable capacity per aircraft given the key network parameters.

{For the second objective, we develop a techno-economic model based on TCO to optimize the deployment cost for both CAPEX and OPEX over a period of time. Given the available bandwidth and the DA2GC link capacity requirement, optimum network deployment with certain number of GSs and number of elements in the antenna arrays is found to minimize TCO. To this end, we presented several TCO-minimizing deployment scenarios based on different cost levels by using genetic algorithms to solve the optimization problem. At the end, the main contribution of this paper is the demonstration of the interplay between different network parameters, i.e., how cost of one parameter changes other network parameters. In this way, future network designers can use our analysis while designing DA2GC networks for continental coverage.}

The remainder of the paper is organized as follows. Section \ref{concept} includes the system model. The beamsteering-loss problem and optimum multifaceted GS design is analyzed in Section \ref{gsarrays}. The multi-user beamforming algorithm and analytical framework is proposed in Section \ref{beamforming}. Section \ref{problem} includes the TCO calculations and the optimization problem. The simulation results for the multi-user beamforming algorithm, GS deployment problem, and TCO optimization are presented in Section \ref{secsim}. Lastly, the conclusions and future work are presented in Section \ref{con}.

\textbf{Notation:} In this paper, we use the following notation: $a$ is a scalar; $\textbf{a}$ is a column vector; $\textbf{A}$ is a matrix; $\textbf{A}^{(k)}$ is the k$^{th}$ column of  $\textbf{A}$; $\vert $\textbf{A}$ \vert$ is the determinant of $\textbf{A}$; $\textbf{A}^{(T,H)}$ is the transpose/conjugate transpose of $\textbf{A}$; $\mathbb{E}[]$ is the expectation operator, $\textbf{I}_N$ represent an identity matrix with dimension $N$. $\mathcal{N}(0,1)$ is a normal distribution with mean $0$ and standard deviation $1$.

\section{System Model}
\label{concept}

This section includes the key performance indicators (KPI) for DA2GC that are taken as baseline for our study. In addition, channel characteristics and main assumptions are discussed. 

\subsection{DA2GC KPIs}
\label{DA2GCKPI}

DA2GC KPIs are proposed by the NGMN Alliance \cite{ngmn}, and these KPIs are presented in Table \ref{reqtab}-A. These estimations are based on $20\%$ active passengers per aircraft and $400$ passengers in each aircraft. Thus, each passenger has $15/(7.5)$ Mbps download/(upload) speeds, so that $1.2/(0.6)$ Gbps download/(upload) speeds are required per aircraft.


For the number of aircraft in one DA2GC cell, NGMN's prediction is $60$ aircraft/$18,000$ km$^2$. Currently, there are $30,000$ daily flights in Europe on average \cite{atag}, and it makes $57$ aircraft/$18,000$ km$^2$ per 24 hours. Therefore, the number of aircraft that are in one cell at the same time will be significantly lower and NGMN's expectations are extremely high. Even at the most extreme case, there are $40$ aircraft near Heathrow airport (the busiest airport in Europe) at the same time. Therefore, we consider $30$ aircraft/$18000$ km$^2$ as the baseline for GS deployment problem.

In addition, the estimated data rate for the NGMN's KPIs are calculated based on 400 passengers on board. This assumption is also unrealistic considering short haul flights that are mostly operated by aircraft with around 200 seats \cite{atag}. The flight statistics for Europe in 2014 are presented in the IATA's annual review for 2015. This report includes the flight statistics for European flights, and the future traffic statistics predicted by using the 3.6\% compound annual growth rate as presented in Table \ref{reqtab}-B. Considering the continental flights will be dominated by mostly short haul flights with 200 passengers and 80\% load factor\footnote{Passenger capacity utilization of aircraft.}, there will be 160 passengers on board on average. With 20\% active aircraft, each aircraft will require $480$ Mbps data rate. For this reason, both $1.2$ Gbps and $480$ Mbps DA2GC link capacities per aircraft are considered in our study while solving the GS deployment problem in Section \ref{secsim}.


\subsection{Channel Characteristics}

DA2GC deployment has dedicated GSs. Therefore, we assume that the deployment is performed to guarantee line-of-sight (LOS) path between GS and aircraft by considering the knowledge of terrain conditions, flight routes and density of traffic. {Therefore, multipath components will not be as strong as the LOS path.} For this reason, we assume that DA2GC channel has a dominant LOS path, and the channel can be modeled as a Rician fading channel. The Rician fading channel is expressed as \cite{tan15}
\begin{eqnarray}
\textbf{H}=\sqrt{{R_K}/({R_K+1})}\bar{\textbf{H}}+\sqrt{{1}/({R_K+1})}\tilde{\textbf{H}},
\end{eqnarray}
where $R_K$ is the Rician factor, $\bar{\textbf{H}}$ and $\tilde{\textbf{H}}$ are the channel gain matrix for LOS and Non-LOS paths, respectively. Non-LOS component is modeled with complex Gaussian matrix such that $\tilde{\textbf{H}}\in \mathbb{C}^{N_R\times N_T}$, where $N_{T,R}$ are the number of elements in the transmitter and receiver, respectively. The LOS component is modeled with the array response vectors such that $\bar{\textbf{H}}=\textbf{a}_R(\theta^{LOS}_R,\phi^{LOS}_R)\textbf{a}^H_T(\theta^{LOS}_T,\phi^{LOS}_T)$, and the array response vectors are expressed as 
\begin{multline}
\textbf{a}_{T,R}(\theta,\phi)=\frac{1}{\sqrt{N}}[1, \hdots, e^{j\frac{2\pi}{\lambda}d(i\sin(\theta)\sin(\phi)+j\cos(\theta))},
\hdots, \\ e^{j\frac{2\pi}{\lambda}d((W-1)\sin(\theta)\sin(\phi)+(H-1)\cos(\theta))}]^T,
\end{multline}
where $i=0,\hdots,W-1$, $j= 0,\hdots,H-1$, $N$ is the number of antenna elements, $d$ is the spacing between array elements, and $\lambda$ is the wavelength. $\theta/\phi$ are the elevation/azimuth angles by assuming the origin is the center of the array, and the zenith direction is vertical to the array. $W$ denotes the number of antenna elements in the azimuth dimension in each row, and $H$ represents the number of antenna elements in the elevation dimension in each column. 

\subsection{Carrier Frequency}

Electronic Communications Committee (ECC) report \cite{ecc} describes the frequency designation discussions and possible regulations for DA2GC. For solving the bandwidth problem, spectrum repurposing/transferring is proposed by ECC. For DA2GC at 5855-5875 MHz and 1900-1920 MHz, there are some regulatory efforts to provide spectrum harmonization in Europe, respectively. However, these bandwidths cannot provide the rates that can compete with SATCOM-based solutions (achieving 70-100 Mbps). For this purpose, spectrum sharing with mobile satellite services (MSS) as complementary ground component and fixed satellite services (FSS) as moving platforms may be promising for DA2GC. Federal Communications Commission (FCC) considers possible frequency sharing between DA2GC and FSS in 14-14.5 GHz \cite{ecc} for US and similar discussion is present in Europe for 17-19 GHz. In this paper, continental coverage for the European Airspace is considered. Therefore, 18 GHz is used to obtain simulation results in this study.

\subsection{MIMO vs. Beamforming}

DA2GC requires improved spectral efficiencies to meet the requirements as outlined in Section \ref{DA2GCKPI}. As discussed in the last subsection, high frequency ranges enable the utilization of large planar antenna arrays in both transmitter and receiver because the dimensions of the arrays are proportional to the wavelength. Large planar antenna arrays can provide beam-tracking capabilities, multi-input multi-output (MIMO) and beamforming gains. 

DA2GC channel is LOS dominated, and has poor scattering environment unlike the rich scattering channels that are often observed in the traditional sub-6GHz wireless communication links. This creates low channel rank, and limits MIMO gains in DA2GC like in SATCOM links. However, dual-polarized antennas can provide additional channel rank to use MIMO with polarization diversity as in SATCOM \cite{dualpol,chang15}. Thus, dual-polarized antenna arrays are promising for DA2GC links. {Therefore, we assume that the MIMO gain in DA2GC is limited only with 2x2 dual-polarization, planar antenna arrays can be utilized to provide array gain via beamforming.} In addition, multi-user beamforming has potential to enhance the utilization of available spectrum by forming separate beam for each aircraft, {such that these aircraft can utilize all of the available spectrum at the same time.}  At the end, this paper focuses on DA2GC with dual-polarized multi-user beamforming capabilities as will be further discussed in Section \ref{beamforming}.

\subsection{Channel Estimation with ADS-B}
\label{adsbsubsec}

Automatic dependent surveillance broadcasting (ADS-B) is developed for ATC and collision avoidance system. This system periodically transmits position of an aircraft based on GPS to all directions. This way, aircraft in the vicinity and ground-based measurement devices collect these signals to monitor trajectory of aircraft. ADS-B based systems are gaining popularity and it has become mandatory for all airlines from 2018 in Europe and 2020 US \cite{ADSB2020}. Therefore, the location information estimated via ADS-B signaling can be exploited by DA2GC GSs as well. 

Compared to the radar-based ATC, which has update times of 6-12 seconds, ADS-B has 0.5-1 second update time. This provides significant improvement in the accuracy of the ATC systems \cite{richards2014}. ADS-B signal can reach up to 200-300 km ranges, and includes position, speed, and altitude of aircraft \cite{kim16}. In \cite{you13}, the comparison between radar and ADS-B based location estimations are presented. According to the results, the error in ADS-B-based location estimations are in the order of a few hundred meters. Thus, the error levels are negligible considering the DA2GC cell range ($20-100$ km). In \cite{nijsure16_1}, the authors propose an ADS-B based multilateration technique using the time-difference-of-arrival (TDOA), angle-of-arrival
(AOA), and frequency-difference-of-arrival (FDOA), and achieve $1.5^\circ$ angular accuracy and $30$ m position accuracy over $100$ nautical miles ($185.2$ km). In \cite{nijsure16_2}, the position estimates from the proposed multilateration system is utilized as an input to beamforming algorithm for drones. 

Beamforming stage can exploit location information through ADS-B because aircraft have specific flight routes and do not make random movements. {Therefore, we assume that the locations of aircraft are determined through ADS-B. In addition, LOS part of the channel gain matrix can be found by just knowing the location of the aircraft since it only depends on the LOS angles. Thus, beam-search and channel estimation do not impose significant problems unlike the traditional terrestrial networks, where users have random mobility and the channel is dominated by multi-path components \cite{emil}. We also model the possible losses that might be caused by the errors in the location estimations in Section \ref{beama}.}

\subsection{Doppler Shift}

Speed of an aircraft at cruising altitudes varies between $800-1000$ km/h. Hence, the Doppler effects at the receiver can be a limiting factor for DA2GC links. However, the characteristics of DA2GC channel alleviates this problem. In the traditional ground cellular networks, there are multipath signals reaching at the receiver following different paths and angles. Therefore, the amount of Doppler shift is different for each path, and this condition creates Doppler spread. Due to the presence of dominant LOS component, DA2GC channels do not have Doppler spread. Instead, spectrum of received signals are shifted by a constant Doppler shift. For this reason, this constant Doppler shift can be pre-calculated and compensated at the receiver.

%

\begin{figure}[!ht]
        \centering
         \begin{subfigure}[b]{0.2\textwidth}
                \centering
                \includegraphics[width=\textwidth]{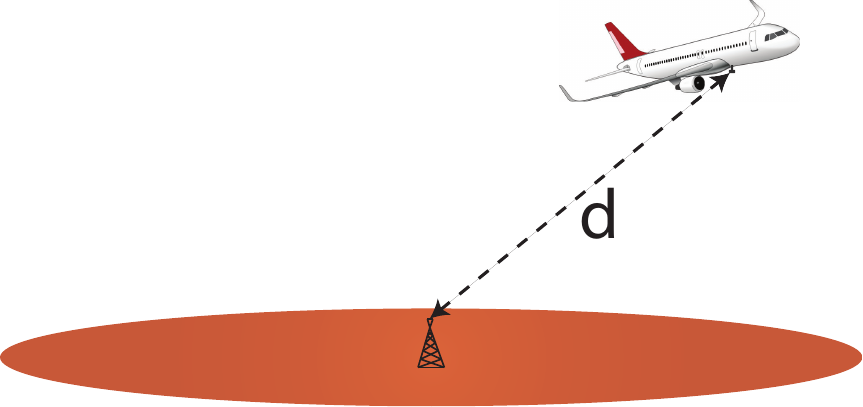}
\caption{DA2GC link.}
                \label{doppler}
        \end{subfigure}
                \begin{subfigure}[b]{0.25\textwidth}
                \centering
                \includegraphics[width=\textwidth]{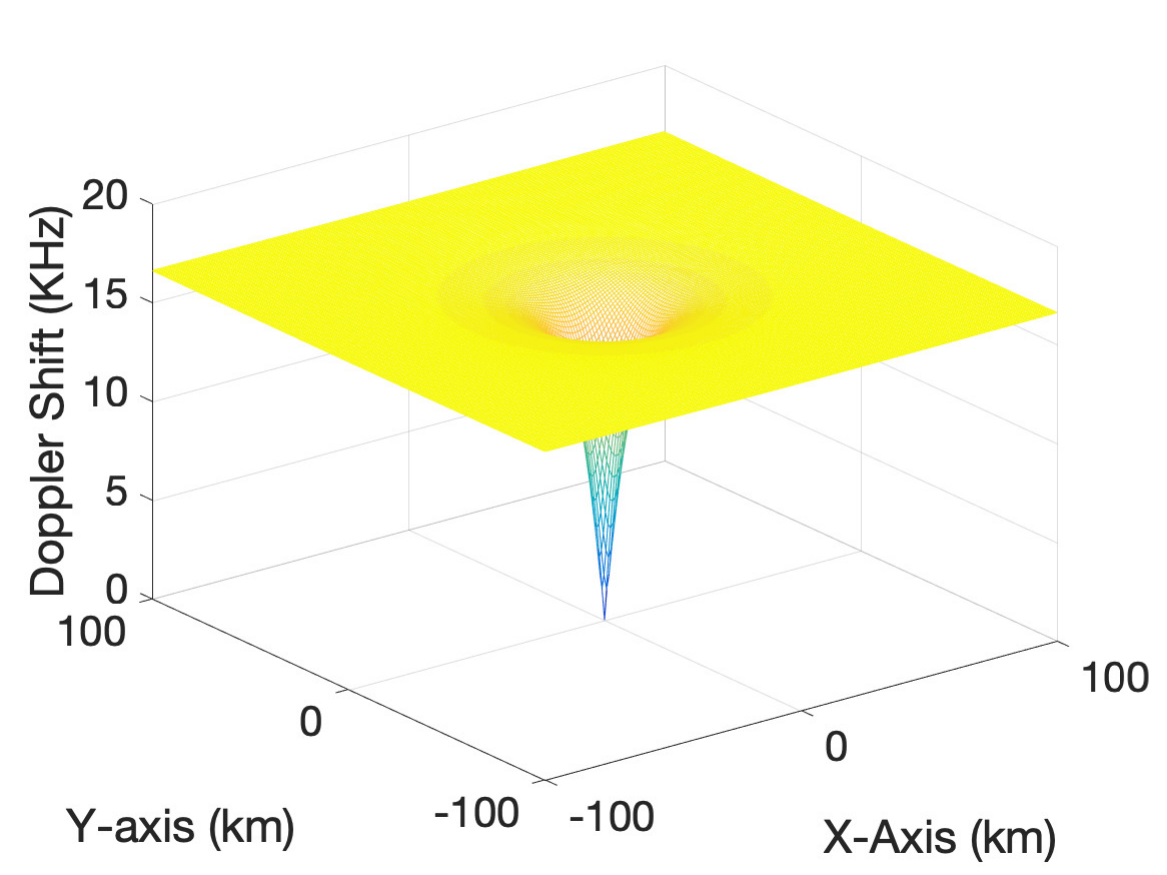}
\caption{Doppler Shift in DA2GC}
\label{dopplers}
        \end{subfigure}
         \caption{(a) DA2GC link, and (b) Doppler Shift in DA2GC at $18$ GHz and $1000$ km/h.}                 \label{dopler_com}
\end{figure}

The link geometry of DA2GC channel is presented in Fig. \ref{dopler_com}(\subref{doppler}). The amount of Doppler shift in the channel is calculated as
\begin{equation}\label{doppeq}
f_D={\dot{d}}/{\lambda},
\end{equation}
where $\dot{d}$ is the rate of change in the DA2GC path length (m/s). According to (\ref{doppeq}), Fig. \ref{dopler_com}(\subref{dopplers}) shows the Doppler shift results for carrier frequency $18$ GHz and aircraft speed $1000$ km/h. As noticed, the highest level of Doppler shift is lower than $16.5$ kHz, and the amount of Doppler shift depends on the radial distance from GS due to angle of the beam. Since the amount of Doppler shift can be estimated by using the location and direction-of-arrival of the received signal, we assume that this shift can be compensated by the receiver by adaptively shifting the spectrum through carrier frequency correction schemes.

\section{Optimal Design of Ground Station Antennas}
\label{gsarrays}

DA2GC GSs utilize antenna arrays to electronically steer multiple beams. However, mutual coupling imposes new constraints in the design of GS antennas. As in Fig. \ref{fig:sys}, the transmission angle from the vertical axis of the cell center to the cell edge aircraft ($\Psi_s$) is very wide due to large cell sizes. For $100$ km and $200$ km inter-site distances (ISDs), the angles from the vertical axis for an aircraft at $9$km altitude are $79.8^\circ$ and $84.9^\circ$, respectively. Even for low ISDs, antenna array needs to scan at least 80$^\circ$ in the elevation, and $360^\circ$ in the azimuth. This condition causes significant decrease in the received power, also known as beamsteering-loss.

\begin{figure}[!ht]
\centering
  \includegraphics[width=0.45\textwidth]{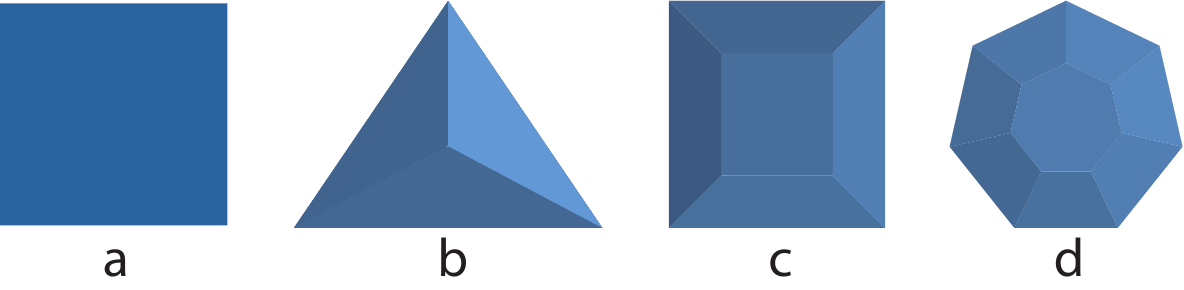}
  \caption{Multifaceted GS antenna array placement (a) $n=1$, $m=0$, (b) $n=2$, $m=3$, (c) $n=3$, $m=4$, (d) $n=3$, $m=7$.}
  \label{figfacet}
\end{figure}

In order to compensate for the beamsteering-loss, multifaceted antenna structure can be utilized as  shown in Fig. \ref{figfacet}. In this way, each antenna array will scan narrower angle span compared to the single-faceted structure, and the amount of loss due to the beamsteering can be kept in minimum levels. {In this paper, an optimization problem is proposed to minimize the total beamsteering-loss in DA2GC GSs. In this way, the optimum number of antenna faces is calculated by utilizing the optimization problem.}

We assume that $n$ is the number of rows that scan the elevation, and $m$ is the number of columns that scan the azimuth in the multifaceted structure. Example figures for different $n$ and $m$ values can be found in Fig. \ref{figfacet}. For odd $n$, one of the antenna array is always placed in the center. This way, the elevation span ($\Psi^S$ depends on ISD) is divided by $n$ as $\Psi^S/n$, and horizontal angle span is divided by $m$ as $\phi^S/m$. At the end, the total number of faces of the structure can be calculated as $(\floor*{n/2}m+\text{mod}(n,2))$, where $\floor*{.}$ is the floor operation. 

\begin{figure}[!ht]
\centering
  \includegraphics[width=0.35\textwidth]{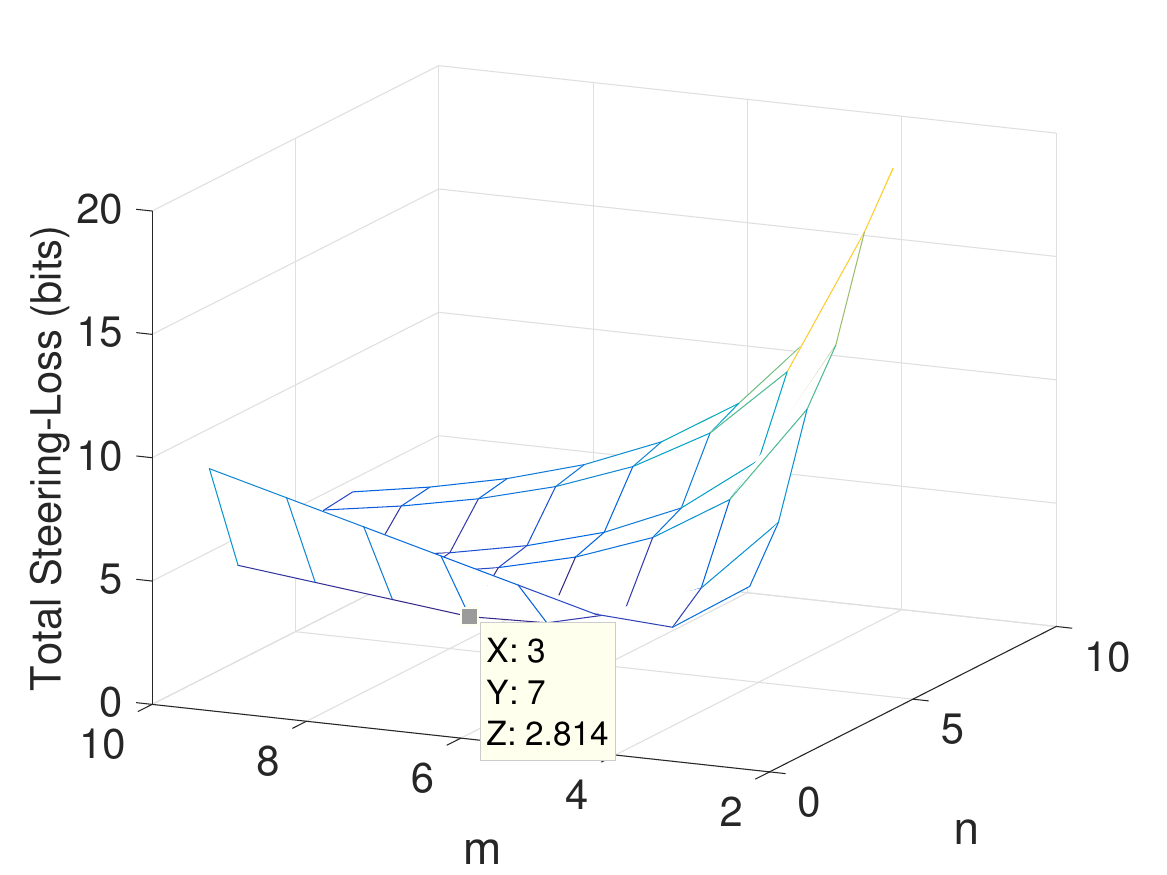}
  \caption{Total beamsteering-loss calculated by (\ref{gsoptpbj}) for ISD=$150$ km.}
  \label{figobj}
\end{figure}

Array gain depends on the cosine of the maximum normal angle from the antenna array \cite{array}. Therefore, gain loss directly affects signal-to-interference plus noise ratio (SINR). By assuming that DA2GC operates at high SINR levels, rate loss due to beamsteering is approximated as \begin{eqnarray}R&=&\log_2(1+\text{SINR}\cos(\max(\Psi^S/n,\phi^S/m))^2), \\
&\approx&\log_2(\text{SINR}\cos(\max(\Psi^S/n,\phi^S/m))^2), \nonumber \\&=&\log_2(\text{SINR})+\log_2(\cos(\max(\Psi^S/n,\phi^S/m))^2).\nonumber
\end{eqnarray}
Therefore, the worst-case rate loss for each antenna array due to beamsteering is determined as $\log_2(\cos(\max(\Psi^S/n,\phi^S/m))^2)$. {Both transmitter and receiver antenna arrays contribute to the beamsteering-loss. That's why, the SINR reduces with cosine square.} The optimal design tries to minimize the total rate loss in the system for all antenna arrays ($(\floor*{n/2}m+\text{mod}(n,2))$). For this purpose, we define an optimization problem to minimize the total worst-case rate loss due to the beamsteering as 
\begin{eqnarray} \label{gsoptpbj}
\underset{n,m}{\text{Minimize}} &&-\log_2(\cos(\max(\Psi^S/n,\phi^S/m))^2)\nonumber\\ && \times(\floor*{n/2}m+\text{mod}(n,2)),\nonumber\\
\text{Constraints:}&&2\leq m, 1\leq n
\end{eqnarray} 
where $\phi^S=\pi$ is the half of the azimuth scan range because the loss depends on the angle with the normal, $\Psi^S=\text{atan}(r_{max}/h_{min})$, where $h_{min}=9$ km is the minimum cruise altitude and $r_{max}$ is the cell range. { $n$ is the number of rows that scan the elevation, and $m$ is the number of columns that scan the azimuth in the multifaceted structure. This optimization problem is solved for fixed $h_{min}$ and $r_{max}$. The objective function for ISD$=150$ km is presented in Fig. \ref{figobj}. The lower bound on $m$ is set as $2$ because $m=1$ can make the objective function go $\infty$. This is a nonlinear integer optimization problem, and the problem is solved by using the genetic algorithms toolbox in MATLAB. The optimal values for the problem are found as $n=3$ and $m=7$ for $150$ km ISD (Figure \ref{figobj}). For these parameters, the optimal structure has $8$ faces in total as presented in Fig. \ref{figfacet}(d). The resulting beamsteering-loss per antenna array is $\zeta=0.35$ bits/channel use for $n=3$ and $m=7$. In addition, the optimal values for other ISDs are presented in Table \ref{ISDtab}. As noticed, the change in the elevation angle range slightly increases with the increasing range, and the number of required faces to scan the azimuth cut decreases with increasing ISD. This is a counterintuitive result, and the reason for this condition is that as the elevation angle range increases, the value of $\Psi^S/n$ in (\ref{gsoptpbj}) increases, but this increase is not enough to make $n$ larger. Therefore, there is no need to use higher values of $m$ as the other term in $\max(.)$ function is dominating. For this reason, we assume $n=3$ and $m=7$ in the rest of the manuscript in order to minimize the beam-steering losses caused by the antenna array.}

\begin{table}[]
\centering
\caption{Optimal Antenna Array Parameters for Different ISDs.}
\label{ISDtab}
\begin{tabular}{|l|c|c|c|}
\hline
\multicolumn{1}{|c|}{\textbf{ISD (km)}} & \textbf{$\phi^S$ ($^\circ$)} & \textbf{$m$} & \textbf{$n$} \\ \hline
$100$                                    & $79.76$                      & $7$          & $3$          \\ \hline
$150$                                    & $83.16$                      & $7$          & $3$          \\ \hline
$200$                                   & $84.85$                      & $6$          & $3$          \\ \hline
$300$                                   & $86.57$                      & $6$          & $3$          \\ \hline
$400$                                   & $87.42$                      & $6$          & $3$          \\ \hline
\end{tabular}
\end{table}


\section{Multi-user Beamforming}
\label{beamforming}

DA2GC systems employ multi-user beamforming techniques to provide increased spectrum efficiency via array gain and increased spectrum utilization by creating a separate beam for each aircraft. In this way, entire spectrum resources can be allocated for each aircraft to enhance the throughput. {For this reason, we tailored the zero-forcing beamforming method for DA2GC.} In addition, we derive an analytical framework to calculate the throughput of a DA2GC GS as a function of antenna array size, number of GSs and bandwidth. 

\subsection{Multi-user Beamforming Model}

In the conventional digital beamforming systems, each antenna element has one radio frequency (RF) chain such that the phase and amplitude of the signal are controlled at the baseband and the resulting signal is up-converted. However, the power consumption of large antenna arrays at high carrier frequencies and high bandwidths will be immensely high if every antenna element has its own RF chain \cite{sp7,sp8,mult}. For this reason, hybrid precoding systems are proposed in which the number of RF chains are limited to provide less power consumption, and we also consider the utilization hybrid precoding systems in DA2GC links. 

\begin{figure}[!ht]
\centering
  \includegraphics[width=0.45\textwidth]{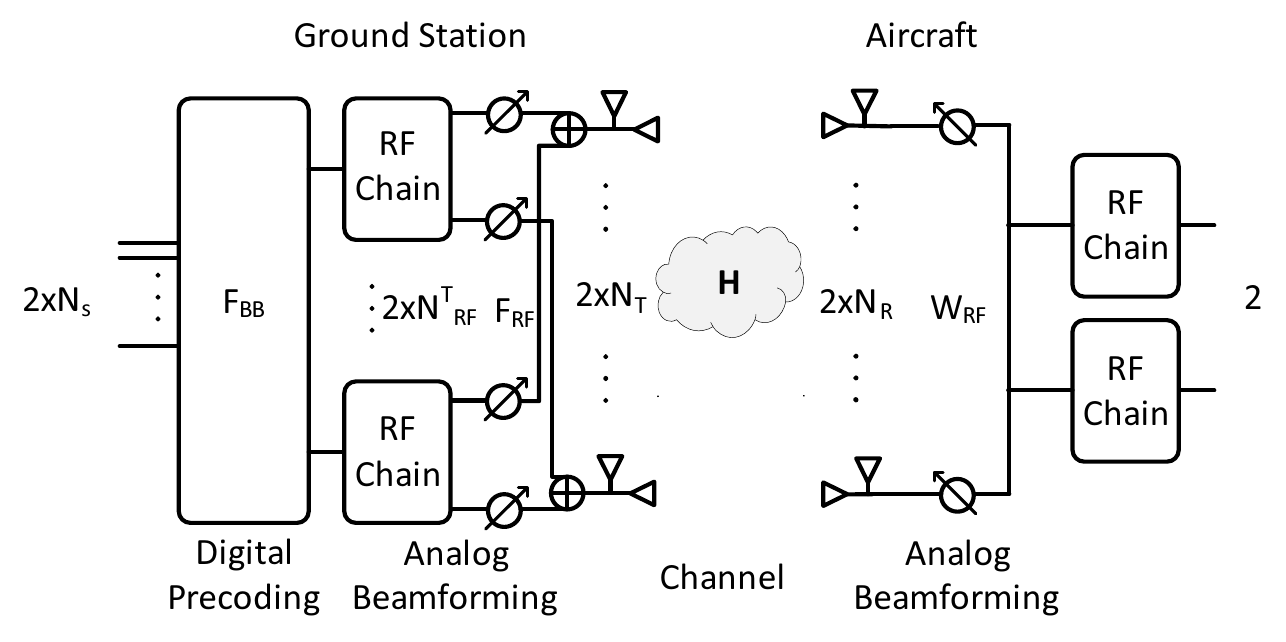}
  \caption{Dual-polarized hybrid multi-user beamforming system model.  }
  \label{sys}
\end{figure}

In Fig. \ref{sys}, the block diagram of dual-polarized hybrid precoding system with $2\times N_T$ number of transmit antennas and $2\times N_R$ number of receive antennas is presented \cite{xia,xia2}. With the hybrid precoding $2\times N_s$ number of data streams can be sent by transmitter chain to receiver chain such that $N_s \leq N^T_{RF} \leq N_T, N_s \leq N^R_{RF} \leq N_R$ where $N^T_{RF}$ and $N^R_{RF}$ are the number of transmitter and receiver RF chains for each polarization, respectively. In this architecture, the transmitter digitally precodes $N_s$ number of multiple data streams to $N^T_{RF}$ number of analog RF chains for each polarization. In this paper, we assume that $N_s>K$ where $K$ is the number of aircraft in a cell. In addition, we assume that polarization diversity provides full diversity gain, thus the proposed algorithm is presented for single polarization. Furthermore, the beamforming matrices for different antenna arrays are calculated together by assuming a single antenna array (as in Fig. \ref{figfacet}(a)) because the actual beamforming vectors can be calculated with rotation matrix. However, the beamsteering-loss is included in the GS deployment calculations.  

The received signal for $k$th aircraft for each polarization can be modeled as 
\begin{eqnarray}
{y}_k
&=&\sqrt{\rho_k}{(\textbf{w}_{RF}^k)}^H \textbf{H}_k \textbf{F}_{RF}\textbf{F}_{BB} {s} + {(\textbf{w}_{RF}^k)}^H\textbf{n}_k,
\end{eqnarray}
where $\rho_k$ is the path-loss for k$^{th}$ aircraft, $s$ is the signal, {$\textbf{H}_k$ is the channel gain matrix for k$^{th}$ aircraft}, $\textbf{F}_{BB}$ is the digital precoding matrix, $\textbf{F}_{RF}$ is the transmitter beamforming matrix, $\textbf{w}_{RF}$ is the receiver beamforming vector, $\textbf{n}_k \sim \mathbb{N}(\textbf{0},\sigma^2 \textbf{I})$, and note that ${(\textbf{w}_{RF}^k)}^H{\textbf{w}_{RF}^k}=1$. 

Therefore, the spectral efficiency of the system with full polarization diversity (channel rank two) can be found as \cite{spencer04}
\begin{eqnarray}\label{actrtotal}
R_{Total}=2\log_{2} \left| \textbf{I}_K+\frac{P_T}{2K N_0 M}\textbf{H}_S\textbf{F}_{BB}\textbf{F}_{BB}^H  \textbf{H}_S^H\right|,
\end{eqnarray}
where $P_T$ is the transmit power, $N_0$ is the noise level ($N_0=-174+\log_{10}(B)$, where $B$ is the bandwidth), and $M$ is the margin for the link. $\textbf{H}_S$ is determined as 
\begin{eqnarray}
{\textbf{H}}_S=\begin{bmatrix}  \sqrt{\rho_k}{(\textbf{w}_{RF}^k)}^H {\textbf{H}}_k \textbf{F}_{RF}\end{bmatrix}_{k\in K\times N^T_{RF}}.
\end{eqnarray} 

Large antenna arrays create narrow beams to provide high array gains. In the traditional terrestrial channels, beam-search and channel estimation are the two major problems for narrow beams. However, these operations can be performed by exploiting location information coming from ADS-B signals as discussed in Section \ref{adsbsubsec}. For this reason, we assume that locations of aircraft are known by GS. In the same way, the locations of the GSs will be known at the aircraft. Therefore, the amount of feedback in the system will be low. 

LOS part of the channel gain matrix can be estimated in the GS by just knowing the relative position of the aircraft with respect to the GS. Since DA2GC links will have large K-factors, we assume that the beamforming process utilizes only LOS part of the channel gain matrix. Thus, DA2GC channel is modeled as $\bar{\textbf{H}}=\textbf{a}_R(\theta^{LOS}_R,\phi^{LOS}_R)\textbf{a}^H_T(\theta^{LOS}_T,\phi^{LOS}_T).$ Since LOS angles are known at the GS, analog beamforming vectors are determined by exploiting the location information such that $\textbf{F}_{RF}^{(k)}=\textbf{a}_T(\theta^{LOS}_{T,k},\phi^{LOS}_{T,k})$ and $\textbf{w}_{RF}^{(k)}=\textbf{a}_R(\theta^{LOS}_{R,k},\phi^{LOS}_{R,k})$. This way, beams are steered directly to aircraft by the closest GS. {This operation does not require any central entity as an AS has its own location and locations of all GSs.}

To reduce the feedback in the system, the digital precoding matrix is determined with the LOS part of the channel gain matrix. This way, transmitter can calculate beamforming matrices without the NLOS part of the channel gain matrix. The estimated channel gain matrix becomes
\begin{eqnarray}
\bar{\textbf{H}}_S=\begin{bmatrix}  \sqrt{\rho_k}{(\textbf{w}_{RF}^k)}^H \bar{\textbf{H}}_k \textbf{F}_{RF}\end{bmatrix}_{k\in K\times N^T_{RF}}.
\end{eqnarray}
The digital precoding matrix is determined with the zero-forcing algorithm as
\begin{eqnarray}\label{fbbbb}
\textbf{F}_{BB}= \bar{\textbf{H}}_S^H ( \bar{\textbf{H}}_S \bar{\textbf{H}}_S^H )^{-1}.
\end{eqnarray}
The main problem in the zero-forcing algorithm is the rank efficiency in $\bar{\textbf{H}}_S$. If two aircraft are located closer than half a beamwidth, the rank of the matrix may not be full. This problem will result in high interference for both aircraft and decrease the performance of the overall system. To avoid this situation, we propose a DFT-based elimination method in the next subsection.

\subsection{DFT-based Elimination}
\label{dftsubsec}
Grassmannian line packing (GLP) system is highly utilized in designing codebooks \cite{kim09}. According to GLP, the optimum codebook design for LOS channels has a uniform angle span, and the angular separation between the codes is given by $2\pi/N$, where $N$ is the number of antenna elements. DFT-based codebooks also have the same angular separation between them. Thus, we develop a DFT-based elimination method instead of comparing all the angles between the aircraft in a cell. DFT-based codebook for planar antenna array is determined as
\begin{multline} 
\textbf{C}^{DFT}_T(l,k)=\frac{1}{\sqrt{N_T}}[1, ..., e^\frac{-j2\pi l}{\sqrt{N_T}},
... , e^\frac{-j2\pi (\sqrt{N_T}-1)l}{\sqrt{N_T}}]^T 
\\ \otimes [1, ..., e^\frac{-j2\pi k}{\sqrt{N_T}},
..., e^\frac{-j2\pi (\sqrt{N_T}-1)k}{\sqrt{N_T}}]^T,\end{multline}
where $\sqrt{N_T}\geq l\geq1$ and $\sqrt{N_T}\geq k\geq1$, and {$\otimes$ is the Hammard product}. In the elimination process, the closest code for each aircraft is selected as
\begin{equation}
c_{k}=\argmax_{p=1,\hdots, N_T} [\textbf{a}^H_T(\theta^{LOS}_{T,k},\phi^{LOS}_{T,k})\textbf{C}_T^{DFT}]^{(p)}.
\end{equation}
After the mapping, if two or more aircraft have the same code, only one of these aircraft is served by the GS. This way, the rank deficiency problem in DA2GC can be eliminated. Assume that the number of active aircraft is represented as $K_{ac}$, the estimated channel gain matrix becomes $\bar{{\textbf{H}}_S}_{_{K_{ac}\times N^T_{RF}}}$, and ${\textbf{F}_{BB}}_{K_{ac}\times K_{ac}}$ is calculated with (\ref{fbbbb}). At the end, the multi-user beamforming algorithm with DFT-based elimination is summarized in Algorithm \ref{pseudocode}. The DFT-based elimination is performed between Line 3-8, and the multi-user zero-forcing beamforming matrices are calculated in Line 9-16.

\begin{algorithm}[!ht]
\footnotesize
\caption{Pseudocode for The Multi-user Beamforming Algorithm}\label{pseudocode}
\begin{algorithmic}[1]
\Procedure{}{}
\\ \textbf{Require} $\theta^{LOS}_{T,k},\phi^{LOS}_{T,k},\theta^{LOS}_{R,k},\phi^{LOS}_{R,k}$ for $k=1,\hdots,K$ \vspace{0.2cm}
\State \% Mapping to DFT vectors 
\For{$k=1,\hdots,K$}
\State $\textbf{m}_T^{(k)}=\displaystyle\argmax_{p=1,\hdots, N_T} [\textbf{a}^H_T(\theta^{LOS}_{T,k},\phi^{LOS}_{T,k}) \textbf{C}^{DFT}_T]^{(p)}$
\EndFor \vspace{0.2cm}
\State \% Eliminate the aircraft on the same code
\State$\textbf{l}_T=\mathrm{unique}(\textbf{m}_T)$ \% Unique is an array function returning unique elements.  \vspace{0.2cm}
\State \% Calculate Analog Beamforming vectors
\For{$k\in \textbf{l}_T$}
\State $\textbf{F}_{RF}^{(k)}=\textbf{a}_T(\theta^{LOS}_{T,k},\phi^{LOS}_{T,k})$
\State $\textbf{w}_{RF}^{(k)}=\textbf{a}_R(\theta^{LOS}_{R,k},\phi^{LOS}_{R,k})$
\EndFor\vspace{0.2cm}
\State \% Estimated Channel Gain Matrix and Zero-forcing beamfoming
\State $\bar{\textbf{H}}_S=\begin{bmatrix}  \sqrt{\rho_k}{(\textbf{w}_{RF}^k)}^H \bar{\textbf{H}}_k \textbf{F}_{RF}\end{bmatrix}_{k\in K\times N^T_{RF}}$
\State $\textbf{F}_{BB}= \bar{\textbf{H}}_S^H ( \bar{\textbf{H}}_S \bar{\textbf{H}}_S^H )^{-1}$
   \EndProcedure
\State \textbf{Return: $\textbf{F}_{RF}, \textbf{w}_{RF}, \textbf{F}_{BB}$ }
\end{algorithmic}
\end{algorithm} 

\subsection{Analytical Model for Cell Throughput}

After the DFT-based elimination, the zero-forcing algorithm keeps the intra-cell interference in negligible levels. Therefore, $R_{Total}$ is given as the spectral efficiency of the system with full polarization diversity and formulated as 
\begin{eqnarray} \label{cellth}
R_{Total}&=&2B\sum_{i=1}^{K_{ac}} \log_2(1+\rho_{ i} \frac{P_T}{2K_{ac}N_0 M}N_T N_R \chi^2),\\
(a)&=&2B\log_2 \left(1+ \frac{P_T}{2K_{ac}N_0 M}N_T N_R\chi^2 \prod_{i=1}^{K_{ac}}\rho_{ i} \right),\nonumber \\
(b)&\approx&2B\log_2 \left(  \frac{P_T}{2K_{ac}N_0 M}N_T N_R\chi^2\prod_{i=1}^{K_{ac}}\rho_{ i}\right)\nonumber \\&=&2B \left( \log_2\left(\prod_{i=1}^{K_{ac}} X\rho_{ i}\right ) \right),\nonumber
\end{eqnarray}
where $B$ is the bandwidth, $X=\frac{P_TN_T N_R\chi^2}{2K_{ac}N_0 M}$ is the constant terms, and $\chi^2$ is the beam-alignment loss. Since the analog beamforming vectors are directly calculated with the exact angles, the full antenna gain is provided by the antenna array ($N_T N_R$) for each active aircraft. {Possible errors in estimating these angles are included in the calculations via the beam-alignment loss.} The approximation between step (a) and (b) is performed by assuming the system operates at high SNR regime. 

The calculations of the expected $R_{Total}$ and $K_{ac}$ are presented in Section \ref{susbsub1} and Section \ref{susbsub2}, respectively. $\chi$ is the loss due to beam-alignment and, the calculation of this parameter is included in Section \ref{beama}.

\subsubsection{\textbf{Path-loss Model}}
\label{susbsub1}

In this paper, the attenuation between aircraft and GS is modeled with the free-space path-loss. Therefore, the expected value for (\ref{cellth}) can be calculated as
\begin{eqnarray} \label{expecrate}
\mathrm{E}\left[R_{total}\right]&=&\mathrm{E}\left[2B\sum_{i=1}^{K_{ac}}\log_2(X\rho_{ i})\right]
 \\&=& 2BK_{ac}\mathrm{E}\left[\log_2(X\rho_{ i})\right]  \leq 2BK_{ac}\log_2\mathrm{E}[X\rho_{ i}]. \nonumber
\end{eqnarray}

The radial distance and altitude of aircraft are modeled with uniform distribution $\mathrm{U}(0,r_{max})$ and $\mathrm{U}(h_{min},h_{max})$, where $r_{max}$ is the maximum cell radius, $h_{min,max}$ are the minimum and maximum altitudes, respectively. Therefore, the expected value for $\mathrm{E}[X\rho_{ i}]$ is calculated as
\begin{multline}\label{rhoest}
\mathrm{E}[X\rho_{ i}]={\frac{(4\pi f)^2X}{c^2}} \int_0^{r_{max}} \int_{h_{min}}^{h_{max}}  {1}/{(r^2+h^2)} \\ \times ({1}/{r_{max}}) {1}/{(h_{max}-h_{min})} dhdr.
\end{multline}

\subsubsection{\textbf{Active Aircraft Number}} 
\label{susbsub2}

\begin{figure}[!ht]
        \centering
         \begin{subfigure}[b]{0.2\textwidth}
                \centering
                \includegraphics[width=\textwidth]{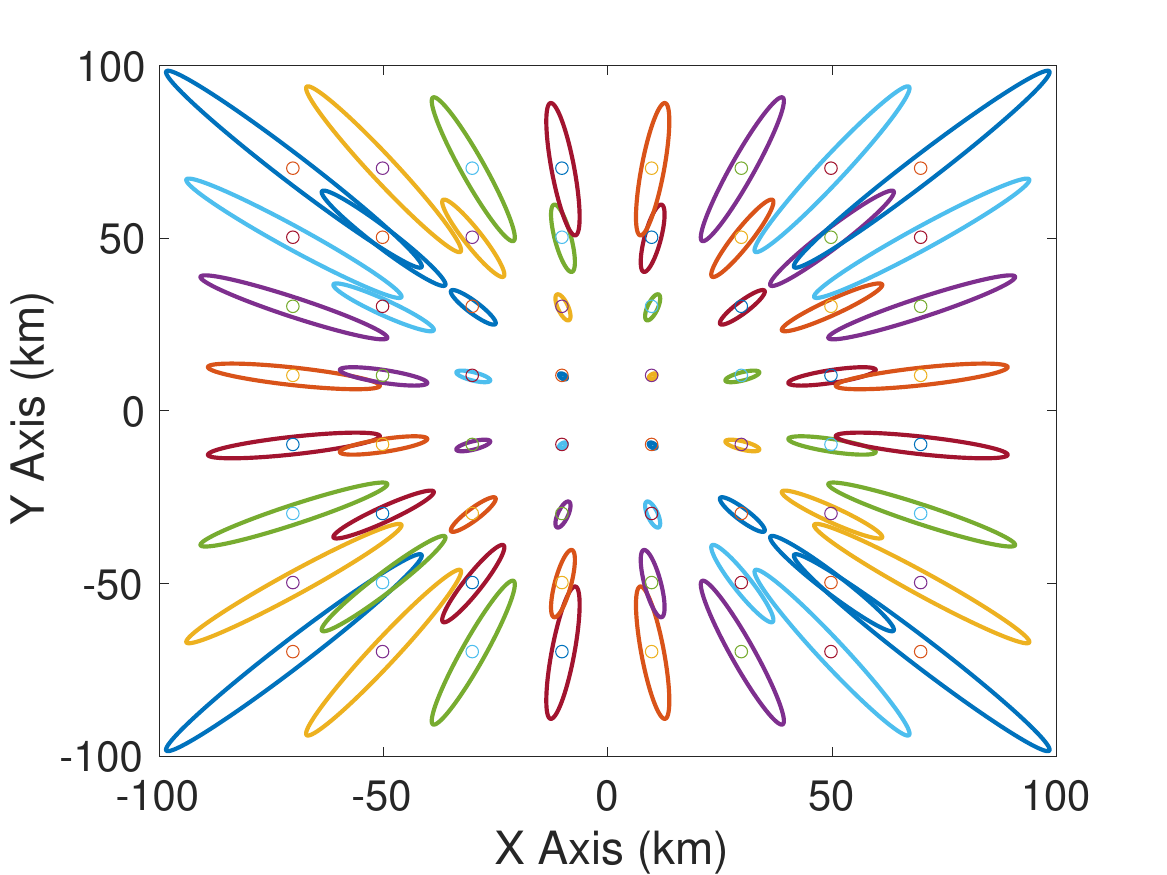}
\caption{{2D-cut of the beams ($z=10$ km.)}}
 \label{visul}
        \end{subfigure}
                \begin{subfigure}[b]{0.25\textwidth}
                \centering
                \includegraphics[width=\textwidth]{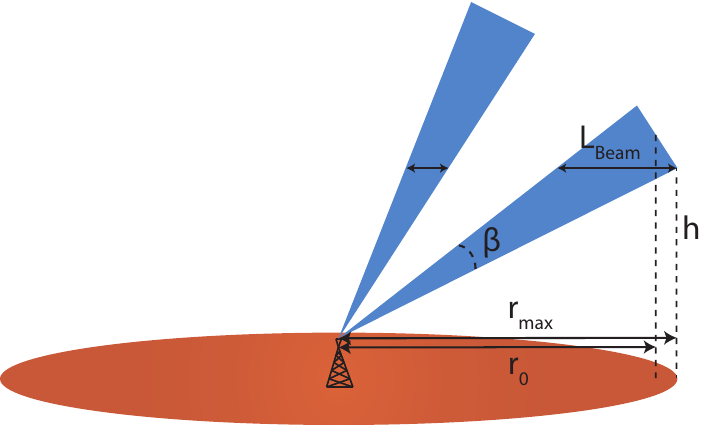}
\caption{Beam length model.}
 \label{beamarea}
        \end{subfigure}
         \caption{Transmission beams from GS.} 
                 \label{ac_visul}
\end{figure}

In the beamforming process, the number of active aircraft are determined based on the number of distinct codes after the DFT codebook mapping as explained in Section \ref{dftsubsec}. For the analytical model, an analytical expression for the number of active aircraft is required to provide tractable expressions. {Beams can be visualized as 3D cones with horizontal and vertical beamwidths (=$101.8/\sqrt{N_{T,R}}$) \cite{array}. Since square antenna arrays are considered in this work, horizontal and vertical beamwidths are equal. More realistic beam models for UAVs are proposed in \cite{qureshi19}. However, we model beams as cones controlled by the beamwidths for determining the number of aircraft that can be supported by a GS. As seen in Fig. \ref{ac_visul}(\subref{visul}) which shows the z-axis cut at $10$ km, beam area increases as radial increase distance between aircraft and GS. Therefore, the probability of being in the same beam increases with cell range.}

%
%

In the analytical model, we assume that each GS has $k$ number of beam choices that can be formed to cover active aircraft. Then, $K$ number of selections ($=$ \# of aircraft in a cell) are performed among $k$ choices with the assumption of each beam choice has the same probability for simplicity. The expected number of distinct beam selections is calculated as 
\begin{equation}\label{findKAC}
\mathrm{E}[K_{ac}]=\left( 1-{(k-1)^K}/{k}^ {K} \right)k,
\end{equation} 
where $k$ is the number of choices that depends on the cell range. For this purpose, we calculate the number of choices with the square of average path length to the cell range as
\begin{multline} \label{kkkkk}
k=\{{L_{Cell}}/{L_{Beam}}\}^2=\{{ r_{0}}/{L_{Beam}}\}^2,  \\
\text{where } L_{Beam}= \int_{0}^{r_{0}} \frac{1}{r_{0} } \frac{h}{ 2}\left\{ \tan\left(\mathrm{atan}(r/h)+{\beta}/{2}\right)  \right. \\ - \left. \tan\left(\mathrm{atan}(r/h)-{\beta}/{2}\right) \right\} dr,
\end{multline}             
where $L_{Cell}$ and $L_{Beam}$ are the diameter of a GS and a beam generated by a GS. $h$ is the aircraft altitude, $r$ is the variable cell range, $\beta$ is the beamwidth. $r_0=
h\tan(\pi/2-\mathrm{atan}(h_{min}/r_{max})-\beta/2)$ is defined to ensure that all created beams by GS will be included in the cell range $r_{max}$ as shown in Fig. \ref{ac_visul}(\subref{beamarea}).

\subsubsection{\textbf{Beam-alignment Loss}}
\label{beama}

In our previous work \cite{dinc17}, it is assumed that the beams from GS and 
AS are perfectly aligned to each other, so that the highest possible array gain is achieved. Considering possible errors in location estimations via ADS-B and beamforming process, there will be a beam-alignment loss. 

In order to model the effects of beam-alignment in the simulations, we assume that the LOS angles will have a Gaussian noise component such that
\begin{eqnarray}
\bar{\theta}^{LOS}_T(\bar{\phi}^{LOS}_T)&=&\theta^{LOS}_T(\phi^{LOS}_T)+\mathcal{N}(0,\Delta),\bar{\theta}^{LOS}_R(\bar{\phi}^{LOS}_R)\nonumber \\ &=&\theta^{LOS}_R(\phi^{LOS}_R)+\mathcal{N}(0,\Delta),\end{eqnarray}
where $\Delta$ is the standard deviation from the exact angles. The beamforming vectors are determined with the modified LOS angles such that $\textbf{F}_{RF}^{(k)}=\textbf{a}_T(\bar{\theta}^{LOS}_{T,k},\bar{\phi}^{LOS}_{T,k})$ and $\textbf{w}_{RF}^{(k)}=\textbf{a}_R(\bar{\theta}^{LOS}_{R,k},\bar{\phi}^{LOS}_{R,k})$. 

For the analytical model, the main lobe of the antenna is modelled with a Gaussian shape $G(\omega)=\exp(-{\omega^2}/{(0.6 \theta_{3dB})^2})$ where $\omega$ is the angle with the normal and $\theta_{3dB}$ is the 3dB beamwidth of the antenna. Therefore, the angular difference between the center of the beams will cause the beam-alignment loss. Assume that the main lobe is shifted in both transmitter and receiver by $\Delta$, the beam-alignment loss {at transmitter or receiever} is calculated as
\begin{eqnarray} \label{chieq}
\chi=\sqrt{\exp\left(-{\Delta^2}/{(0.6 \theta^T_{3dB})^2}\right)}\sqrt{\exp\left(-{\Delta^2}/{(0.6 \theta^R_{3dB})^2}\right)},
\end{eqnarray}
where $\theta^{(T,R)}_{3dB}$ are the 3dB beamwidth for transmitter and receiver, and are calculated for planar antenna arrays as ($101.8/\sqrt{N_{T,R}}$) \cite{array}. {The total beam-alignment loss is calculated as $\chi^2$}.

\subsubsection{\textbf{Analytical Throughput Estimation}} At the end, the expected DA2GC cell throughput can be estimated by combining (\ref{expecrate}) and (\ref{rhoest}) as
\begin{multline} \label{finalest}
R^{est}_{total} \approx 2BK_{ac}\left[\log_2  \left({\frac{(4\pi f)^2}{c^2}}\frac{P_TN_T N_R\chi^2}{2K_{ac}N_0 M} \right. \right. \\ \left. \left. \int_0^{r_{max}} \int_{h_{min}}^{h_{max}}  \frac{1}{r_{max}(r^2+h^2)(h_{max}-h_{min})} dhdr\right) -\zeta\right], 
\end{multline}
where $K_{ac}$ is calculated via (\ref{findKAC}) and (\ref{kkkkk}), {beam-alignment loss at an antenna array} $\chi$ is determined via (\ref{chieq}), and $\zeta$ is the beamsteering-loss as discussed in Section \ref{gsarrays}.

\subsection{Intercell Interference}

\begin{figure*}[!ht]
        \centering
        \begin{subfigure}[b]{0.35\textwidth}
                \centering
                \includegraphics[width=\textwidth]{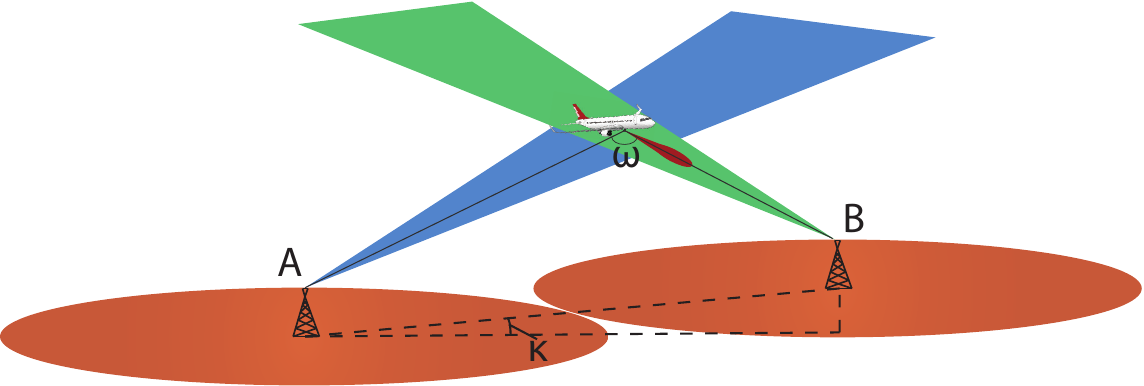}
\caption{Intercell-interference model.}
\label{figaligninter}
        \end{subfigure}
         \begin{subfigure}[b]{0.29\textwidth}
                \centering
                \includegraphics[width=\textwidth]{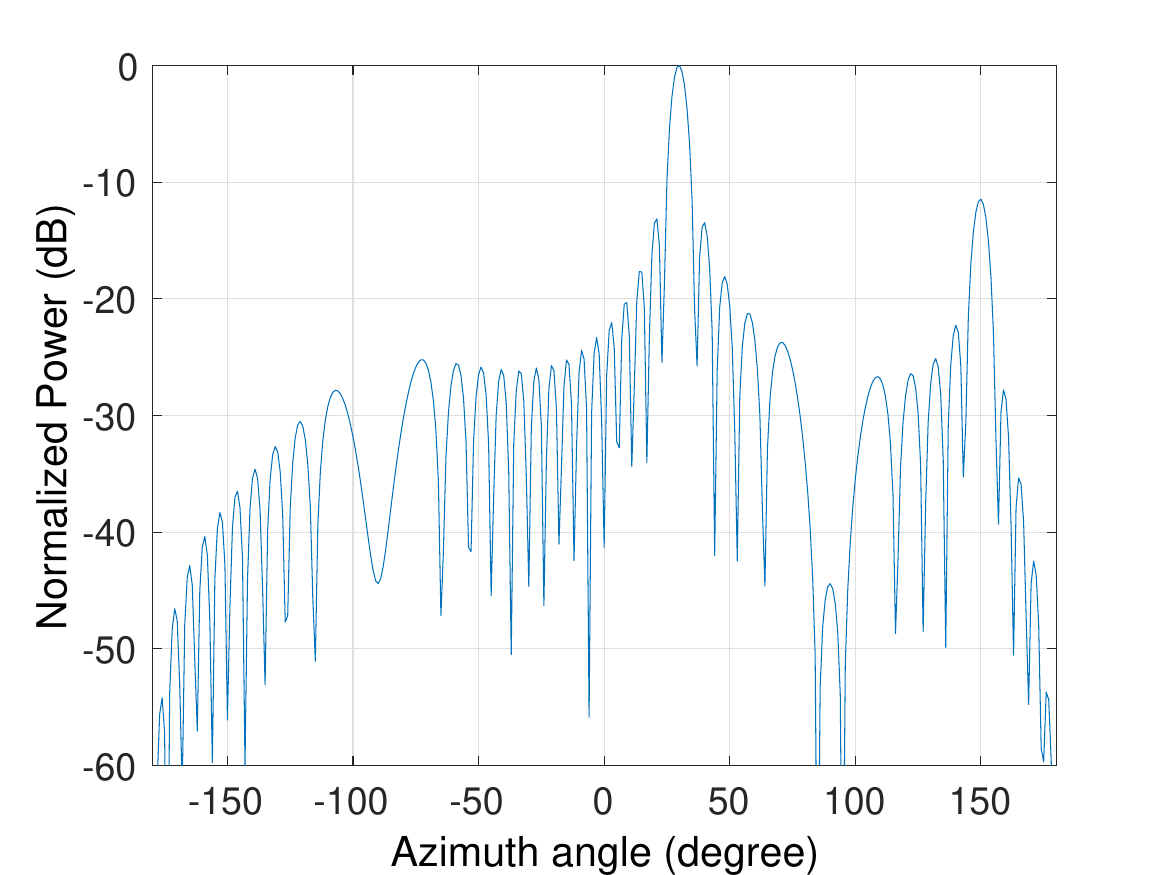}
\caption{Azimuth cut.}
\label{intAZ}
        \end{subfigure}
                \begin{subfigure}[b]{0.29\textwidth}
                \centering
                \includegraphics[width=\textwidth]{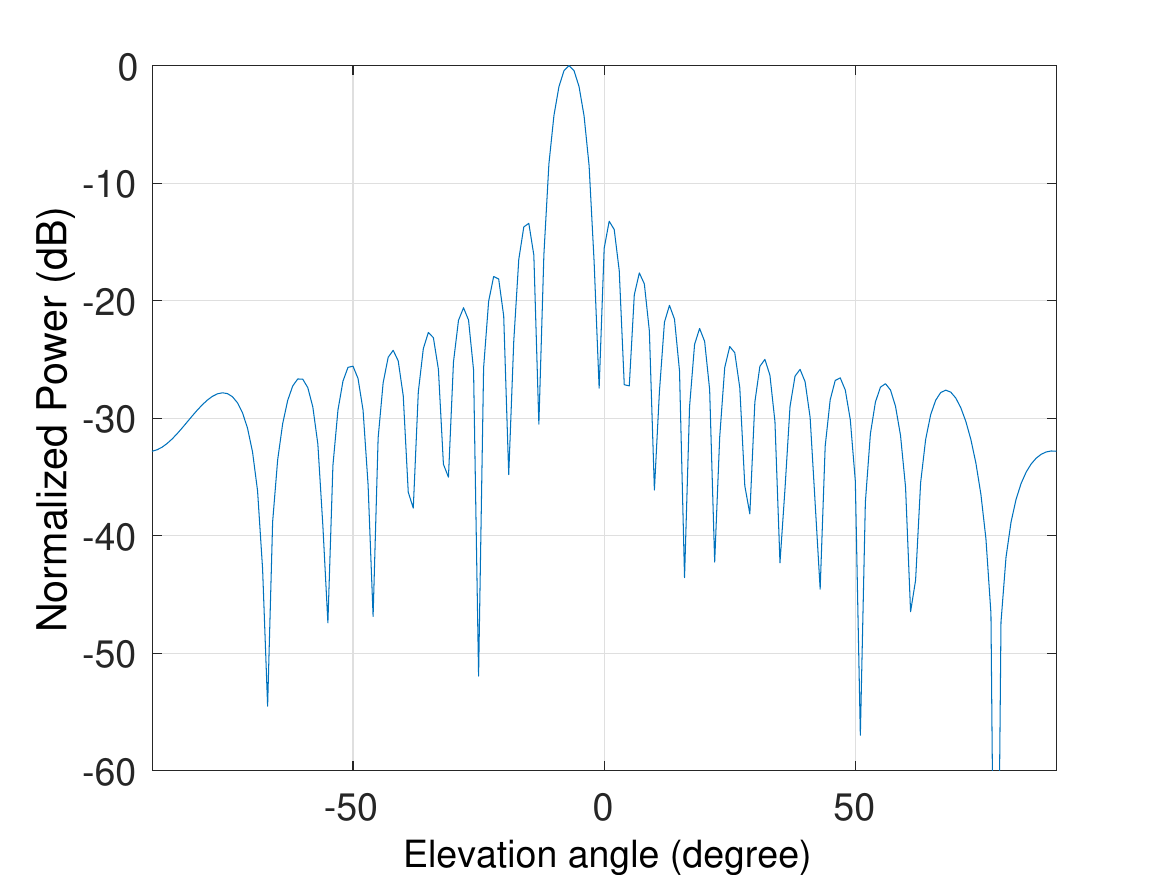}
\caption{Elevation cut.}
\label{intEL}
        \end{subfigure}
        \caption{(a) Intercell-interference model, (b, c) array response for the aircraft located between two GS.} 
        \label{intercellarray}
\end{figure*}

In the derivation of the analytical expression, intercell interference is neglected, and this section includes the justification for this assumption. In the proposed multi-user beamforming algorithm, both transmitter and receiver have antenna arrays, and receivers form their beams to the closest GS. This way, the interference coming from the neighboring cells will have high angle ($\omega$) with respect to the mainlobe of the receiver as presented in Fig. \ref{intercellarray}(\subref{figaligninter}). Therefore, this angular separation will provide the aircraft in the middle to have negligible intercell interference. 

Assume an example scenario as presented in Fig. \ref{intercellarray}(\subref{figaligninter}), the two cells are located such that they have angular separation $\kappa=30^\circ$. This way, the intended GS (B) is in the azimuth angle of $30^\circ$, and the interfering GS (A) is in $-150^\circ$.  Both GSs are in the same elevation angle with respect to the aircraft. Fig. \ref{intercellarray}(\subref{intAZ},\subref{intEL}) show the array response of the aircraft beamformed towards the GS B. As noticed, the mainlobe is located at $30^\circ$ whereas the difference between the mainlobe and the interference direction ($-150^\circ$) is around $-30$ dB. {These results clearly show that the effect of intercell interference between DA2GC GSs is negligible.} For this reason, the analytical expressions do not include the effect of intercell interference since it can be avoided through beamforming in the aircraft terminal. {In addition, ASs will have the exact location of all GSs, and GSs can estimate the position of ASs through ADS-B as the considered cell ranges ($50-150$ km) are lower than the range of ADS-B signals ($200-300$ km) \cite{kim16}}.


\section{Ground Station Deployment Problem}
\label{problem}

This section provides the GS deployment to minimize the TCO and determine the feasible points for a set of network design parameters. The analytical model is utilized to calculate the GS specifications to achieve certain data rate values by considering two DA2GC backhaul link capacity requirement scenarios: $1.2$ Gbps and $480$ Mbps as discussed in Section \ref{DA2GCKPI}. The following subsections include the TCO model, mathematical definitions of the objective function, i.e., TCO model, and the constraints, i.e., backhaul capacity, power limitations, antenna size. 

\subsection{TCO Model}

\begin{figure}[!ht]
\centering
  \includegraphics[width=0.35\textwidth]{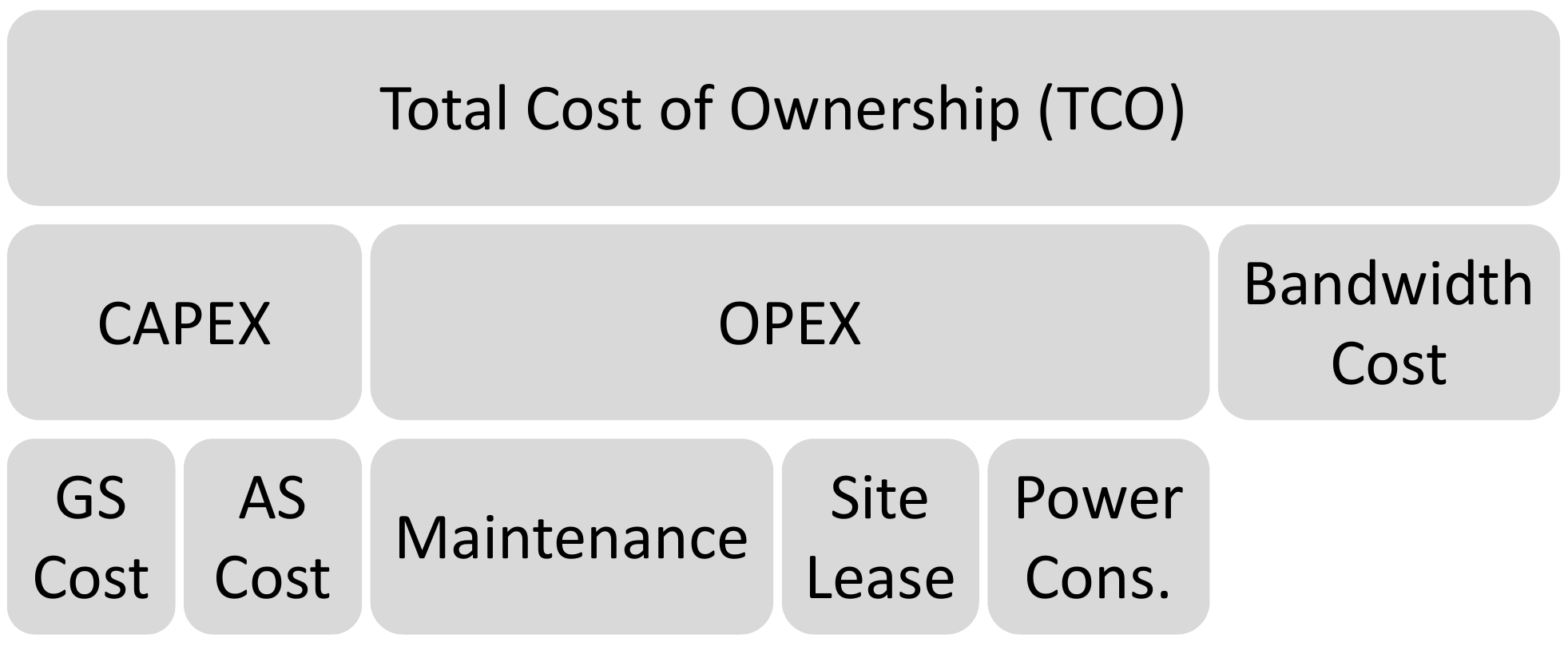}
  \caption{TCO breakdown.}
  \label{tcobreak}
\end{figure}

We utilize a simplified TCO model for 10-year of operation time such that some of the costs, that are not changing (or slightly changing) based on the network parameters are omitted. Backhaul cost, for example, is considered as a fixed cost because the total aggregate backhaul capacity of the DA2GC network is not scale with the same range as the number of GSs. In addition, we assume that GSs are deployed in the existing cell towers, where fiber infrastructure and grid connectivity are already available. Hence, backhaul cost is not considered in this model. At the end, the TCO model considers CAPEX, OPEX and bandwidth cost as in Fig. \ref{tcobreak}. In CAPEX, the contributing costs are GS cost and air station (AS) cost. In OPEX, the following costs are included in the proposed TCO model: power consumption, site lease, maintenance.

\subsubsection{\textbf{CAPEX}} DA2GC CAPEX is modeled with the cost of GS and AS. CAPEX for the GS and AS ($C_{GS}$ and $C_{AS}$, respectively) includes the cost of the baseband processing unit and antenna arrays. As discussed in Section \ref{gsarrays}, each side of communication will have 8 antenna arrays to compensate for the beamsteering-loss. Since the cost information for such antenna arrays are not available in the literature, we calculate the cost of GS and AS based on variable price per antenna element. In \cite{ibmanten}, the authors present an antenna array having 64 elements with beamforming capabilities, and this antenna is built by combining four antenna arrays, each having 16 antenna elements. Considering this, we assume that the total cost of antenna array is linearly scales with the number of antenna elements. 

For the cost of baseband processing units, we use the cost of regular LTE base stations ($C_{base}$=10K\euro{}) \cite{jan14,jan142}. At the end, the total CAPEX of the system is modeled as 
\begin{eqnarray}\label{capex_1}
C_{CAPEX}&=&C_{GS} N_{GS}+C_{AS} N_{AS}\nonumber \\ &=&(C_{base} +8N_T  C_{element} )N_{GS} \\&&+(C_{base} +8N_R  C_{element}) N_{AS}.
\end{eqnarray}
where $C_{element}$ is the CAPEX of each antenna element, $N_{GS}$ and $N_{AS}$ are the numbers of GSs and ASs. The number of GS is calculated based on dividing the European area to the cell area as $N_{GS}=A_{Europe}/(\pi r_{max}^2)$, where $A_{Europe}=10,180,000$ km$^2$. The total number of passenger aircraft in Europe is $8,142$ as presented in Table \ref{reqtab}-B; thus, we calculate the TCO by assuming $5,000$ of them will have DA2GC capabilities.

\subsubsection{\textbf{OPEX}} OPEX cost for the service includes the site lease, power consumption and maintenance costs. According to \cite{tower}, the site lease in Europe varies between $900-1700$ \euro{}/month, so we use average site lease of 1300 \euro{}/month=$C_{Lease}$. The yearly maintenance cost $C_{Main}$ is assumed as 10\% of the CAPEX as proposed in \cite{jan3}. 

For the power model, two states of the GS are considered: idle state and transmitting state. In the idle mode, the power consumption is just limited to $P_F=118.7$ W as also introduced in the techno-economic studies \cite{jan14,jan142}. The power model for the transmitting state is taken from a recent paper focusing on the energy efficiency in hybrid precoding systems \cite{hybridRF,emil15}. Therefore, the total power consumption in Wh of the GS can be calculated as 
\begin{equation}
P_{BS}=P_F T_{10}+(P_{T}/\eta+N_{RF} P_{RF}+P_{syn})T_{T},
\end{equation}
where the power consumption of RF chains is $P_{RF}=1W$, the frequency synthesizer power consumption is $P_{syn}=2W$, $\eta$ represent the power amplifier efficiency that is assumed as 22\% \cite{paeeff}. Since GSs will be always on, the power of the idle state multiplies by the number of hours in 10-year operation time $T_{10}=24\times365\times10$ whereas the transmitting state time is represented as $T_{T}$. {In this model, GS is assumed to always be on (idle state).}

For calculating the transmitting state times, we use the real-world statistics presented in Table \ref{reqtab}-B. In this table, 2014 is taken from the real statistics from \cite{atag}, and the estimations for 2020 and 2030 is calculated based on the $3.6\%=\tau$ yearly growth rate in the air transportation industry (the growth rate is specified in the report \cite{atag} as well.). Therefore, the total number of daily flight hours $T_{Total}=9,347,619$ is shared among all GSs such that the total flight hours inside one cell is given by $T_{CellAgg}=T_{Total}/N_{GS}$. However, a GS can support up to $K_{ac}$ aircraft and $T_{CellAgg}$ includes the sum of all flight hours. Since a GS will not always have the maximum number of aircraft, the average 10-year transmitting state time of a cell is calculated as
\begin{eqnarray}\label{ttdenklem}
T_{T}={T_{Total}}/{K_{average}N_{GS}} \times 365,
\end{eqnarray}
where $K_{average}$ is calculated as $K_{ac}/2$. 

The average kWh cost of electricity is $0.12\euro{}$/kWh$=c_{kWh}$ in Europe \cite{kwhprice}, and this cost is represented in the equations as the Wh cost of electricity $c_{Wh}=c_{kWh}/10^3$ Therefore, the power consumption for 10-year can be calculated as 
\begin{multline} \label{cpppp}
C_{Power}=c_{Wh} N_{GS} P_F  T_{10}  +c_{Wh} N_{GS} \\ \times [P_{T}/\eta + N_{RF} P_{RF}+P_{syn}] T_{T}  \sum_{i=0}^9 (1+\tau)^i,
\end{multline}
where the effect of increasing air traffic is included with a yearly growth rate of $\tau=3.6\%$. {The power calculations are performed with the average operating times of GSs via (\ref{ttdenklem}).}

At the end, the OPEX cost can be represented as 
\begin{eqnarray}\label{copex_1}
C_{OPEX} &=& C_{Lease}+C_{Main}+C_{Power},
\end{eqnarray}
where the 10-year tower lease is $C_{Lease}=1.3 K\euro{} \times 12 \times 10$, the 10-year maintenance cost is $C_{Main}=C_{CAPEX}\times0.9$, and $C_{Power}$ is calculated with (\ref{cpppp}).

\subsubsection{\textbf{Bandwidth}}
In \cite{janspec}, the authors review the spectrum prices for many European countries. We also use the methodology used in \cite{janspec}. The cost of spectrum is determined based on per MHz per population price. As the population, the total number of passengers in Europe is taken as the baseline which is $\approx$ 1,006 billion$=N_{pop}$ as in Table \ref{reqtab}-B. The spectrum price for the $2.6$ GHz in Europe is around $0.01-0.3$\euro{}/MHz/Pop, {and the unit of the bandwidth cost is given in per MHz per population}. {Similar spectrum prices are seen for 5G at 3.4 GHz, and cellular operators has paid around $0.1-0.15$\euro{}/MHz/Pop in the United Kingdom \cite{kavanagh18}.} Since DA2GC frequencies can be also utilized for the ground terrestrial network, the cost of the spectrum will be lower compared to the LTE{/5G} frequencies. Therefore, the TCO calculations are performed with varying spectrum prices ($=c_B$) between $0.001-0.01$\euro{}/MHz/Pop. Thus, the cost of the spectrum is calculated as 
\begin{equation}\label{CB_1}
C_{B}=c_B B N_{pop}.
\end{equation}

\subsection{Constraints and Optimization Problem}
\label{susecprob}

To optimize TCO, the optimization problem can be formulated as
\begin{eqnarray} 
 &&\underset{r_{max},N_T,N_R,P_T,B}{\text{minimize}} C_{CAPEX}+C_{OPEX}+C_{B} \label{tcofinaleq}\\ 
&&\text{Constraints:} \text{  }R_{Th}\leq (R^{est}_{Total})/K , \label{c11}\\
 && \text{  } 10\log(P_T/\text{1 mW})\leq 60 \text{dBm},\label{c22} \\
 && \text{  } \sqrt{N_T}<2L_T/\lambda, \sqrt{N_R}<2L_R/\lambda \label{c33}
\end{eqnarray}
where $C_{CAPEX}$, $C_{OPEX}$, $C_{B}$ are defined in (\ref{capex_1}), (\ref{copex_1}), (\ref{CB_1}), respectively. {(\ref{c11}) is to guarantee that the estimated average rate of an aircraft is higher than the threshold value ($R_{th}$).} The total transmit power budget of a GS is limited to $60$ dBm as in (\ref{c22}). The dimensions of the GS and AS is assumed to be limited by $L_T=0.5$ m and $L_R=0.25$ m as in (\ref{c33}), respectively. 

\section{Simulation Results}
\label{secsim}

This section includes the simulation results for the beam-alignment loss, multi-user beamforming with DFT-based elimination, GS deployment and TCO optimization. {The results presented in Subsection A-B are included to compare the analytical model with Monte Carlo simulations to show the accuracy of the proposed analytical framework.}

\begin{figure*}[!ht]
        \centering
        \begin{subfigure}[b]{0.32\textwidth}
                \centering
                \includegraphics[width=\textwidth]{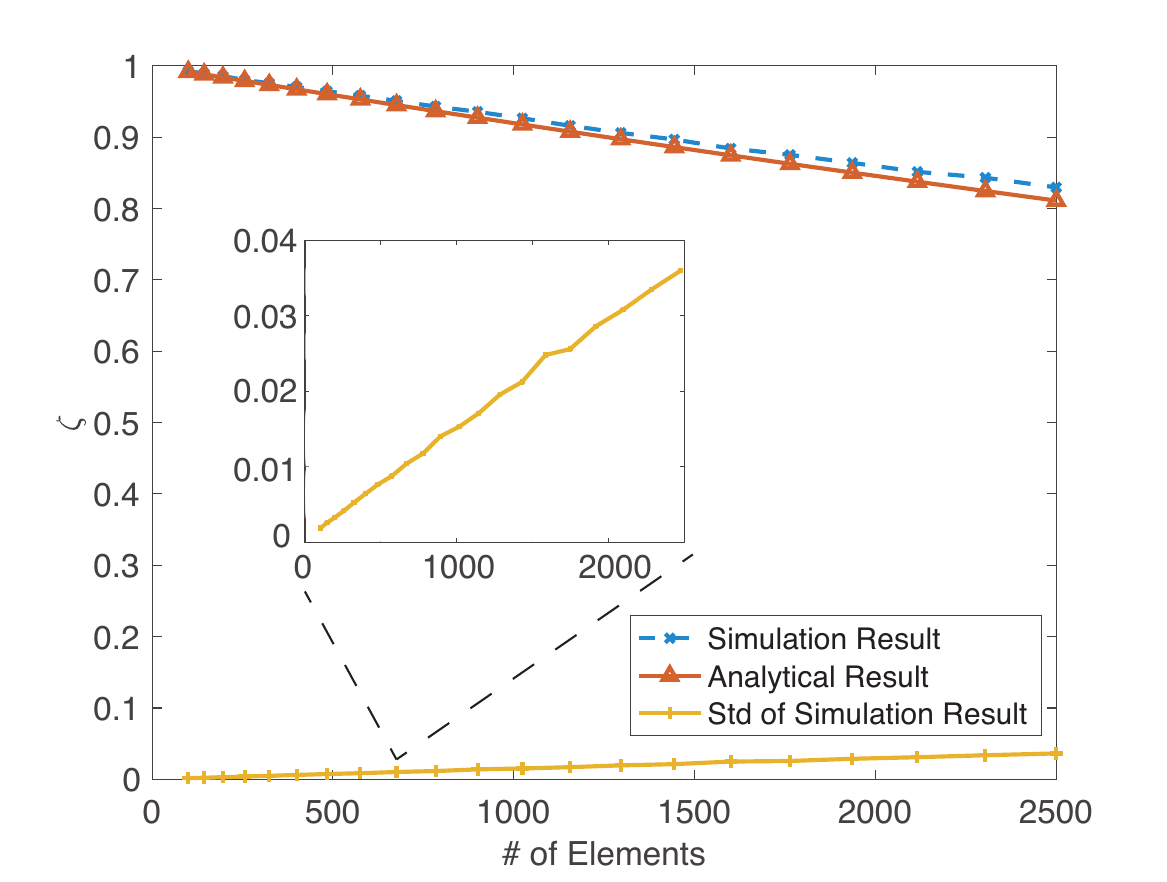}
\caption{Beam-alignment loss.}
        \end{subfigure}
         \begin{subfigure}[b]{0.32\textwidth}
                \centering
                \includegraphics[width=\textwidth]{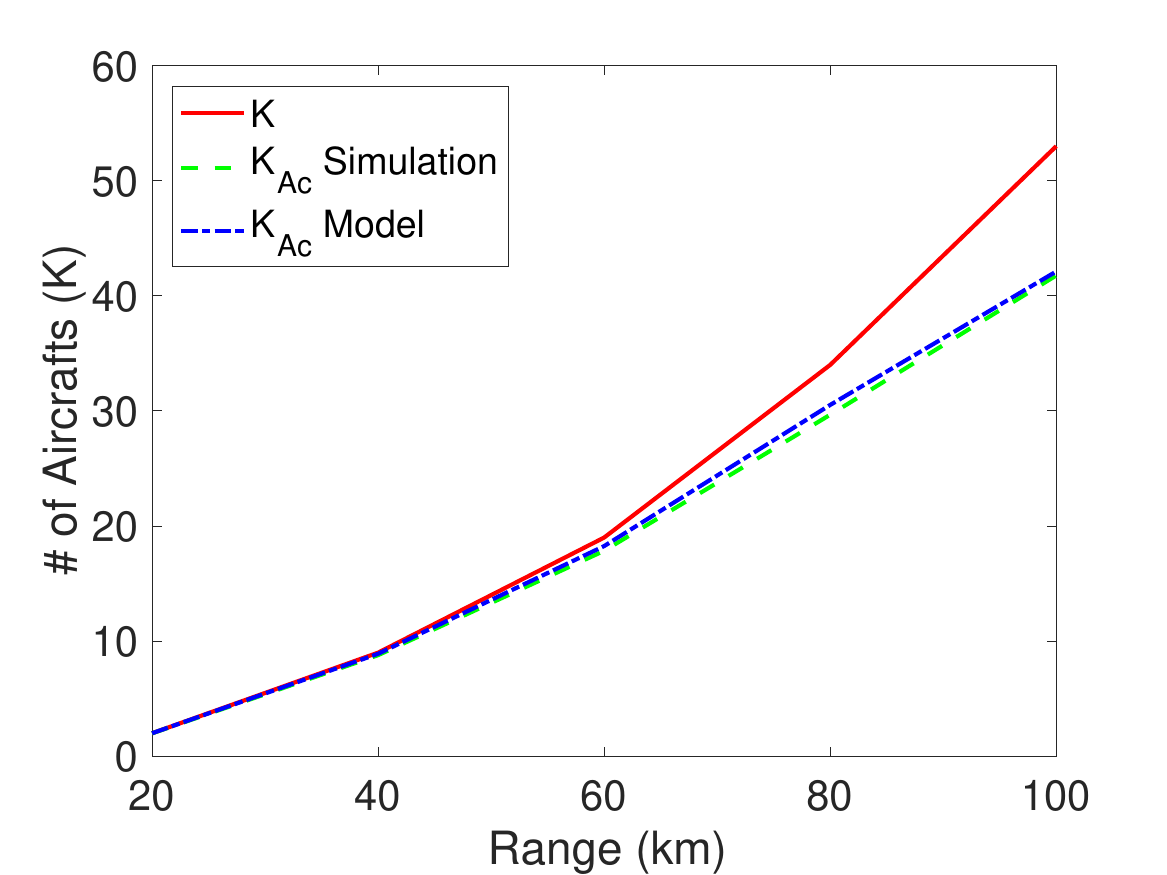}
\caption{$N_T=625$}
        \end{subfigure}
                \begin{subfigure}[b]{0.32\textwidth}
                \centering
                \includegraphics[width=\textwidth]{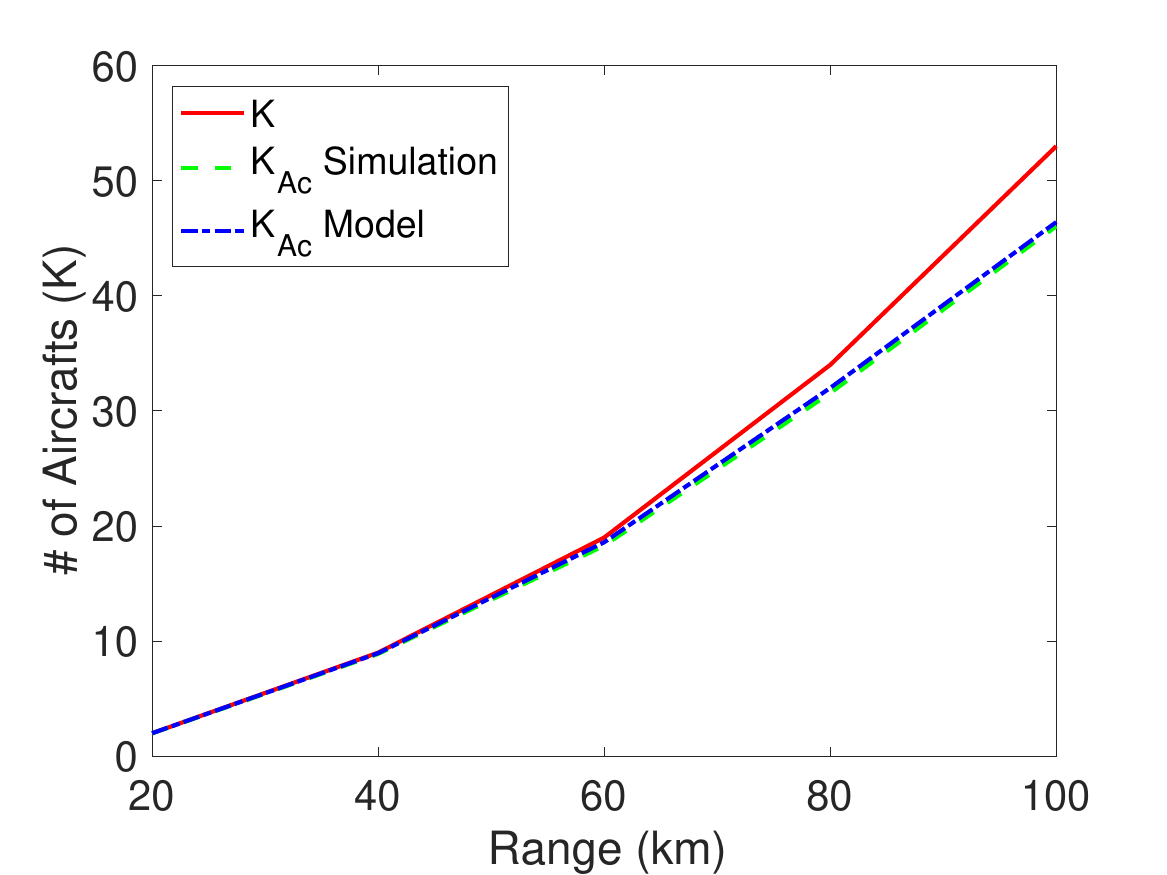}
\caption{$N_T=1225$}
        \end{subfigure}
         \caption{(a) Beam-alignment loss for $\Delta=0.5^\circ$, $r_{max}=75$ km, $h_{min}=9$ km, and (b,c) the results for the total number of aircraft and the number of active aircraft for $N_R=400$ and, $30$ Aircraft/$18000$km$^2$.} 
                 \label{ac}
\end{figure*}

\begin{figure*}[!ht]
        \centering
         \begin{subfigure}[b]{0.32\textwidth}
                \centering
                \includegraphics[width=\textwidth]{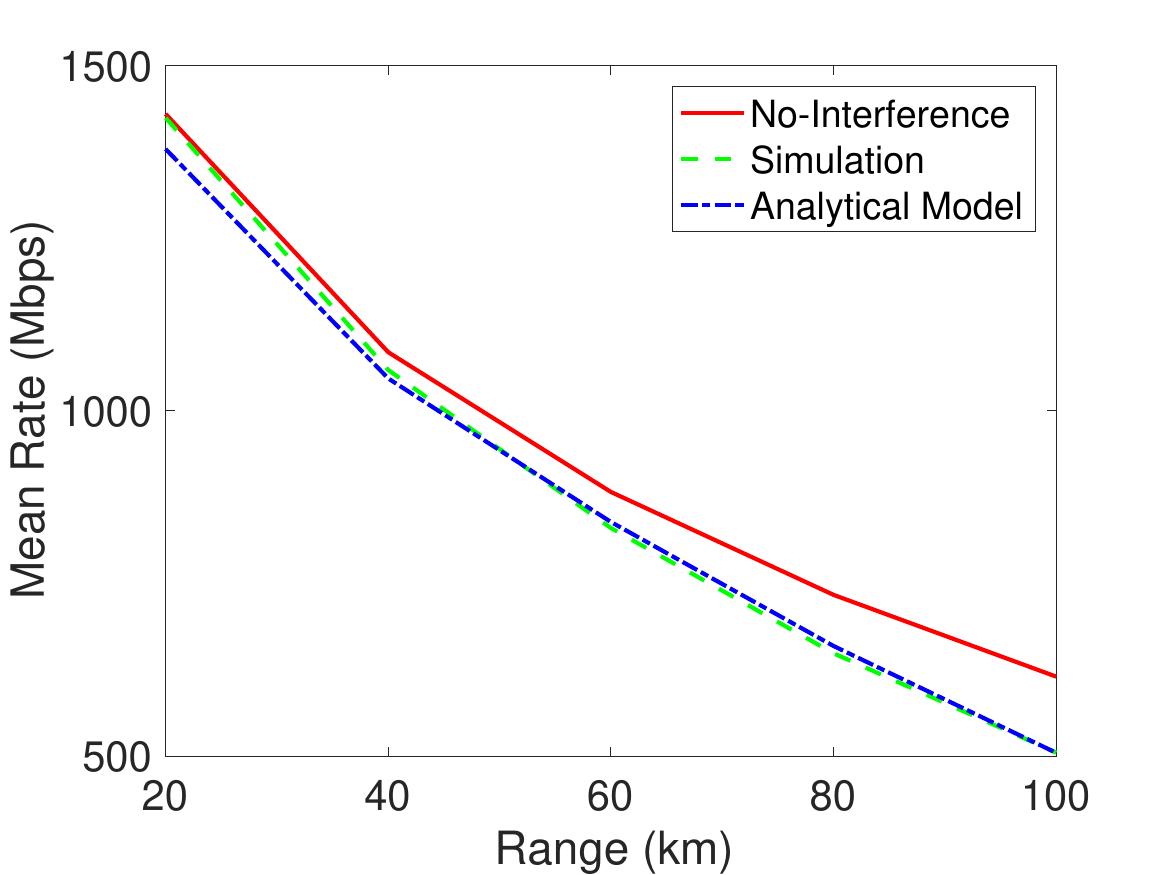}
\caption{Mean rate $N_T=625$}
        \end{subfigure}
                \begin{subfigure}[b]{0.32\textwidth}
                \centering
                \includegraphics[width=\textwidth]{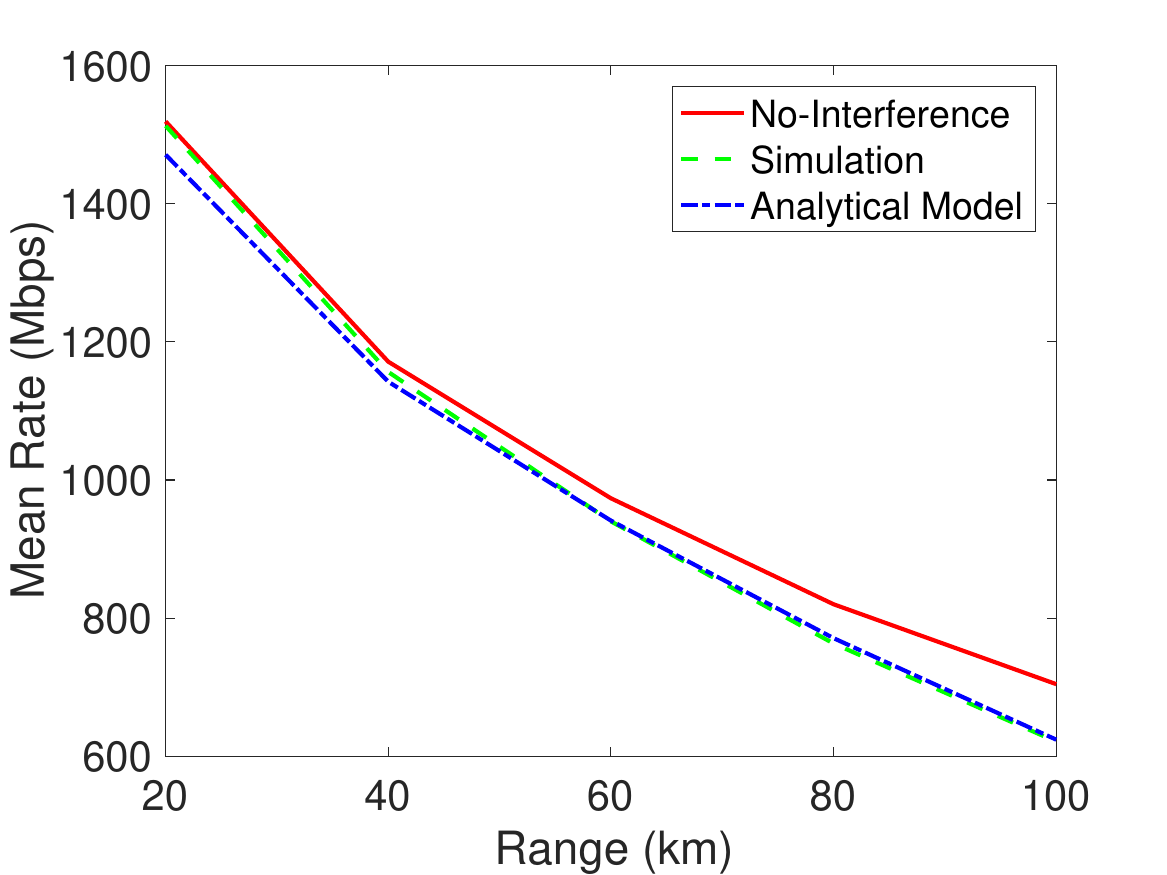}
\caption{Mean rate for $N_T=1225$}
        \end{subfigure}
            \begin{subfigure}[b]{0.32\textwidth}
                \centering
                \includegraphics[width=\textwidth]{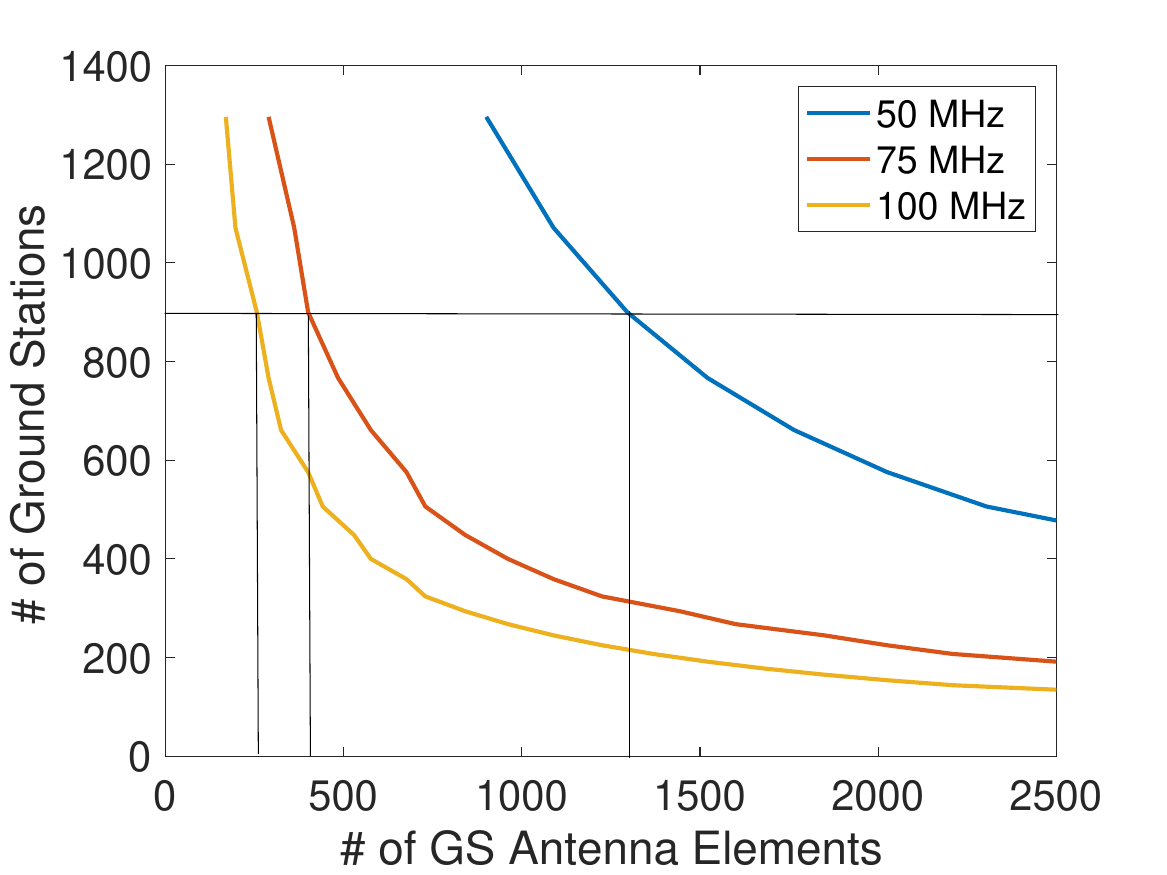}
\caption{\# of GSs vs. array size.}
        \end{subfigure}
        \caption{(a,b) The results for the mean rate of DA2GC for $N_R=400$ and, $30$ Aircraft/$18000$km$^2$, and (c) trade-off between \# of GSs and array size.} 
        \label{rate}
\end{figure*}

\subsection{Beam-alignment Loss}

Fig. \ref{ac}(a) presents the comparison of simulation results with the analytical model in terms of beam-alignment loss as a function of antenna elements. The simulation results are calculated with 
\begin{multline}\chi_{sim}=[\textbf{a}_T(\bar{\theta}_T^{LOS},\bar{\phi}_T^{LOS})^H\textbf{a}_T({\theta_T}^{LOS},{\phi_T}^{LOS})  \\ \textbf{a}_R(\bar{\theta}_R^{LOS},\bar{\phi}_R^{LOS})^H\textbf{a}_R({\theta_R}^{LOS},{\phi_R}^{LOS})],\end{multline} where ${\phi_{T,R}}^{LOS}$ are distributed randomly over $[0,2\pi)$ and ${\theta_{T,R}}^{LOS}$ are distributed uniformly over $[0,\text{atan}(h_{min}/r_{max}))$. The standard deviation of the simulation results is also presented in Fig. \ref{ac}(a) considering average of 10000 realizations. As noticed, the standard deviation of the simulation results is increasing with increasing number of antenna elements because the effect of beam-alignment loss becomes higher for lower beamwidth values. As noticed, the analytical model has very close behavior with the simulation results. The gap between analytical and simulation model increases with the number of antenna elements; however, the loss is overestimated by the analytical model and the difference is lower than $3\%$.


\subsection{Simulation Results For Analytical Framework}
\label{simm}

In this subsection, we present the simulation results for the proposed multi-user beamforming method and the results for the analytical method. The simulations are performed in MATLAB. Each result is generated with $1000$ realizations. The main aim of this section is to show that the proposed analytical framework is consistent with the simulation results. In the simulations, we use the following simulation parameters: carrier frequency $18$ GHz, bandwidth $50$ MHz, transmit power $45$ dBm, $20$ dB K-factor, $M=10$ dB link margin, $N_R=400$, and $30$ Aircraft/$18000$km$^2$. The aircraft are placed randomly in the cruising altitudes $9-13$ km.

\begin{figure*}[!ht]
        \centering
         \begin{subfigure}[b]{0.19\textwidth}
                \centering
                \includegraphics[width=\textwidth]{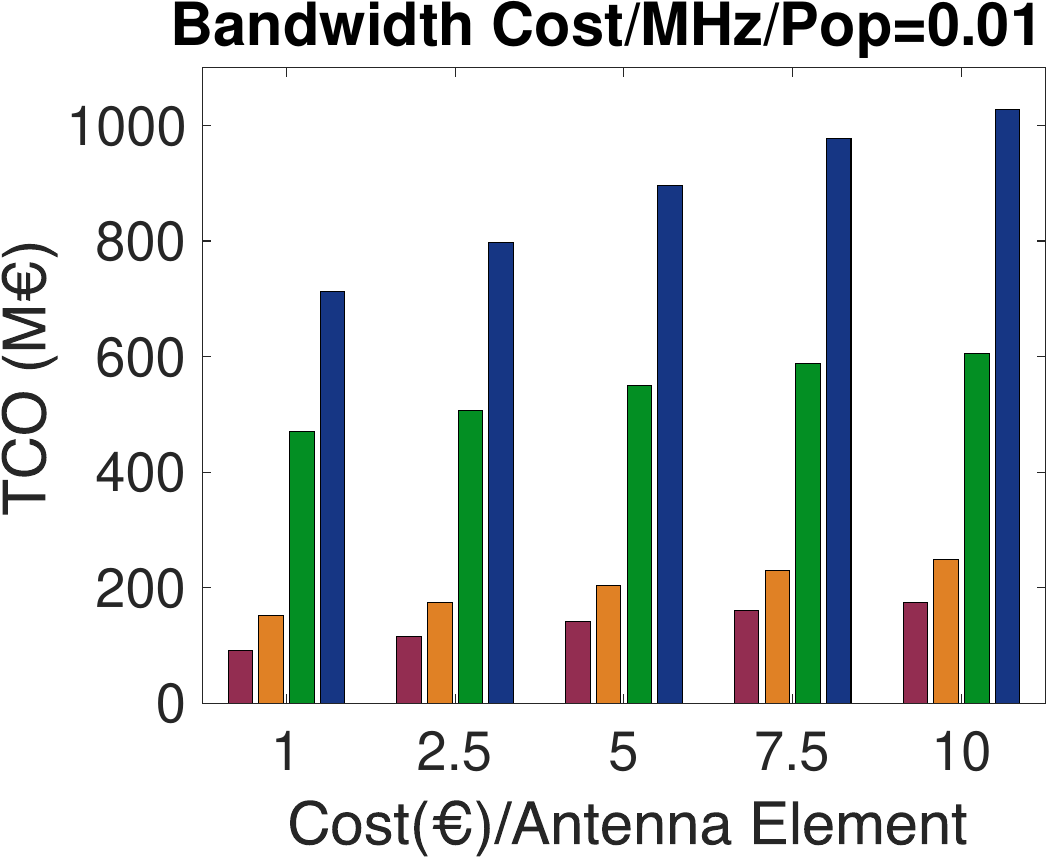}
        \end{subfigure}
        \begin{subfigure}[b]{0.19\textwidth}
                \centering
                \includegraphics[width=\textwidth]{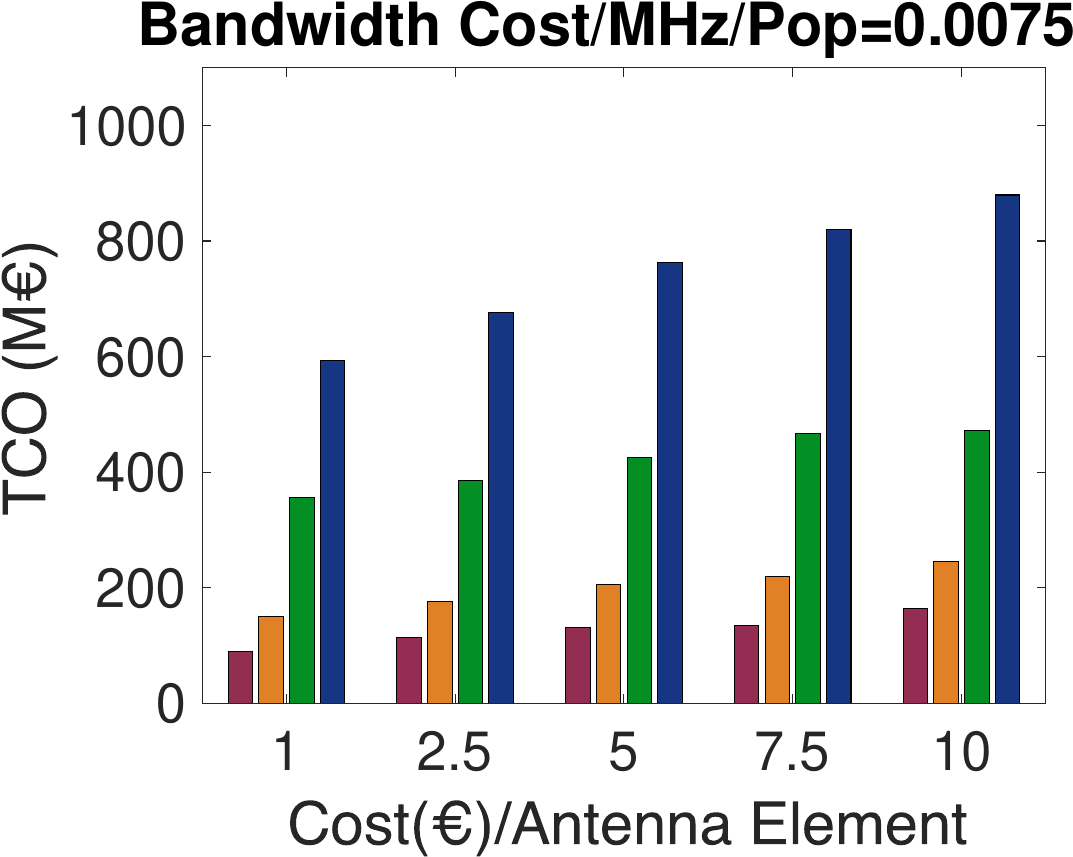}
        \end{subfigure}
        \begin{subfigure}[b]{0.19\textwidth}
                \centering
                \includegraphics[width=\textwidth]{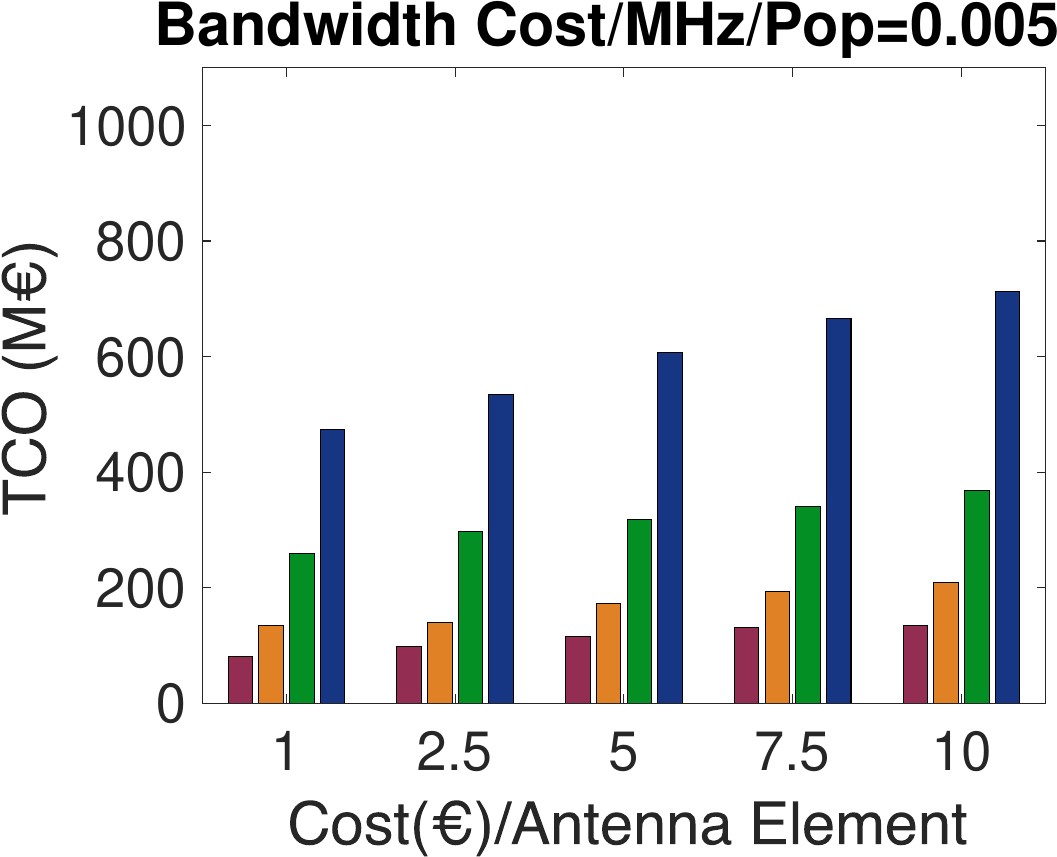}
        \end{subfigure}
        \begin{subfigure}[b]{0.19\textwidth}
                \centering
                \includegraphics[width=\textwidth]{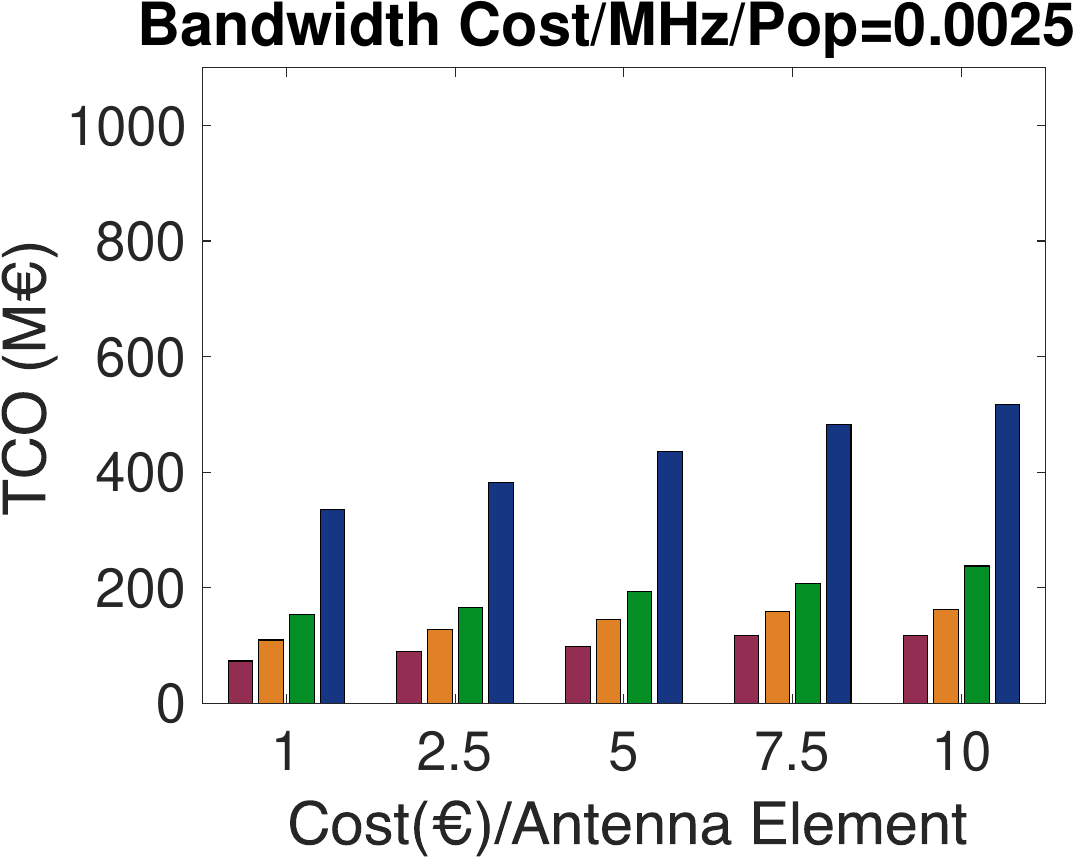}
        \end{subfigure}
        \begin{subfigure}[b]{0.19\textwidth}
                \centering
                \includegraphics[width=\textwidth]{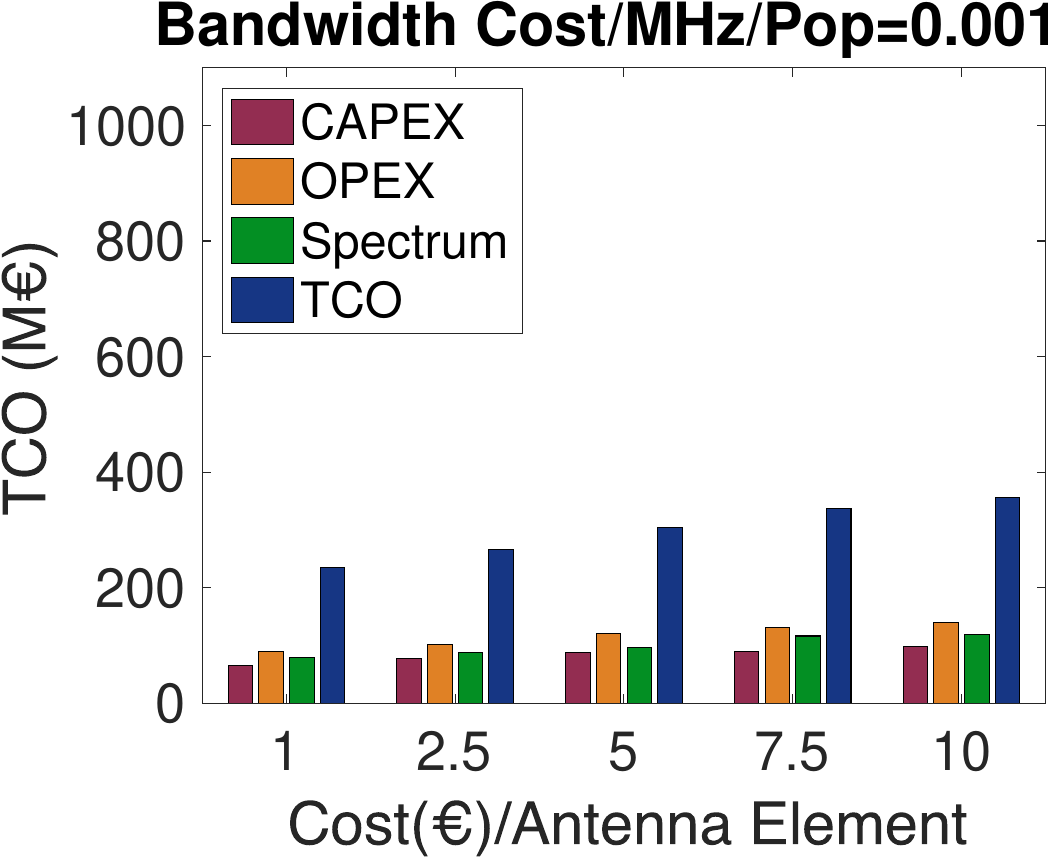}
        \end{subfigure}
        \caption{Distribution of TCO with different cost types for different antenna element and bandwidth costs to achieve $1.2$ Gbps.}
        \label{TCObar12}
\end{figure*}

\begin{figure*}[!ht]
        \centering
         \begin{subfigure}[b]{0.19\textwidth}
                \centering
                \includegraphics[width=\textwidth]{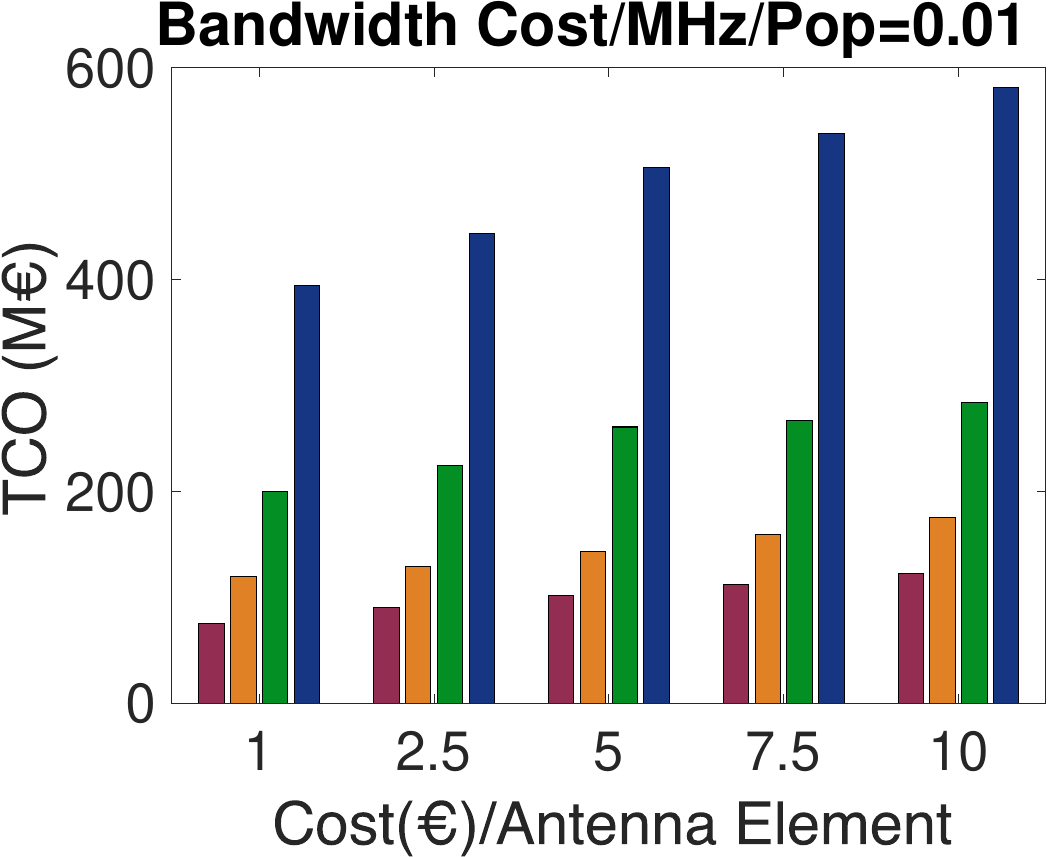}
        \end{subfigure}
        \begin{subfigure}[b]{0.19\textwidth}
                \centering
                \includegraphics[width=\textwidth]{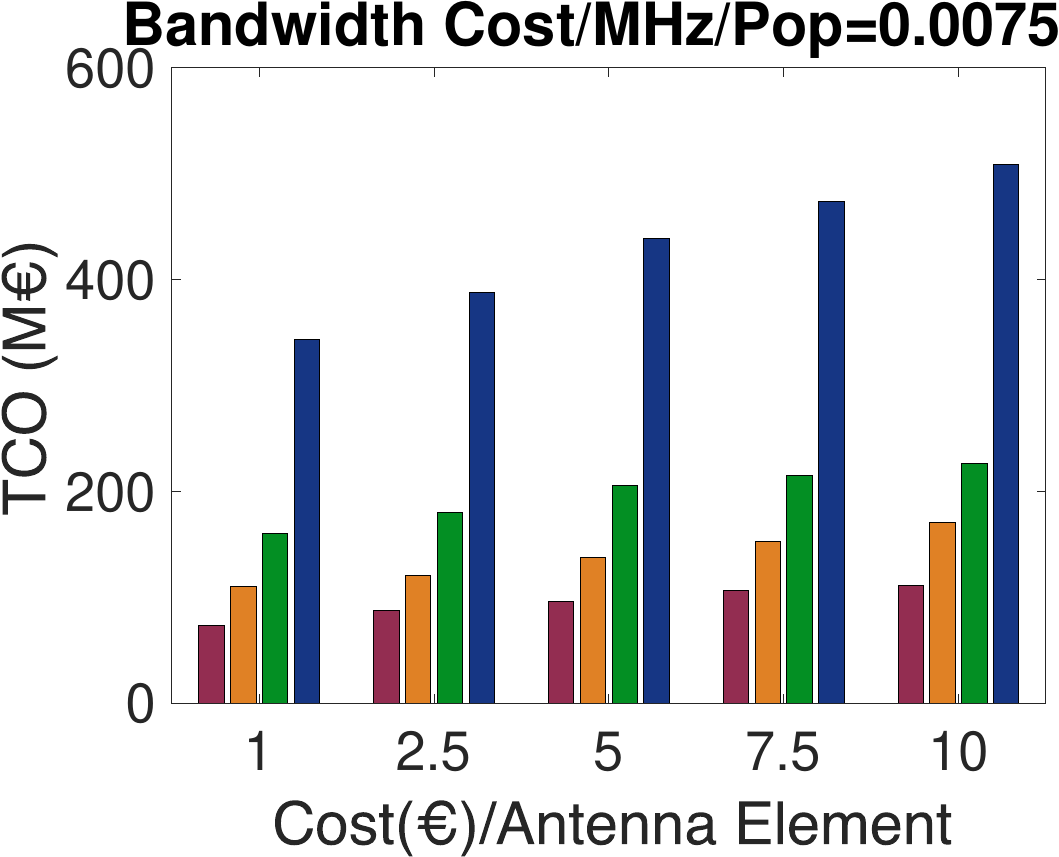}
        \end{subfigure}
        \begin{subfigure}[b]{0.19\textwidth}
                \centering
                \includegraphics[width=\textwidth]{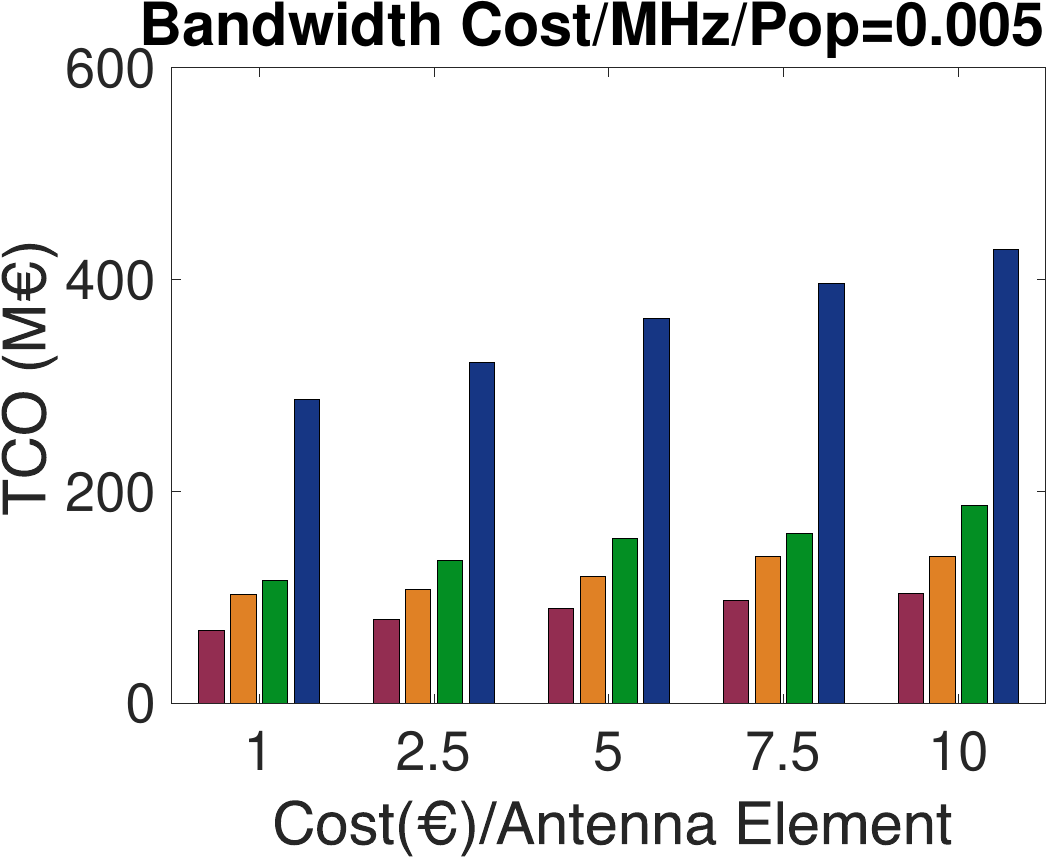}
        \end{subfigure}
        \begin{subfigure}[b]{0.19\textwidth}
                \centering
                \includegraphics[width=\textwidth]{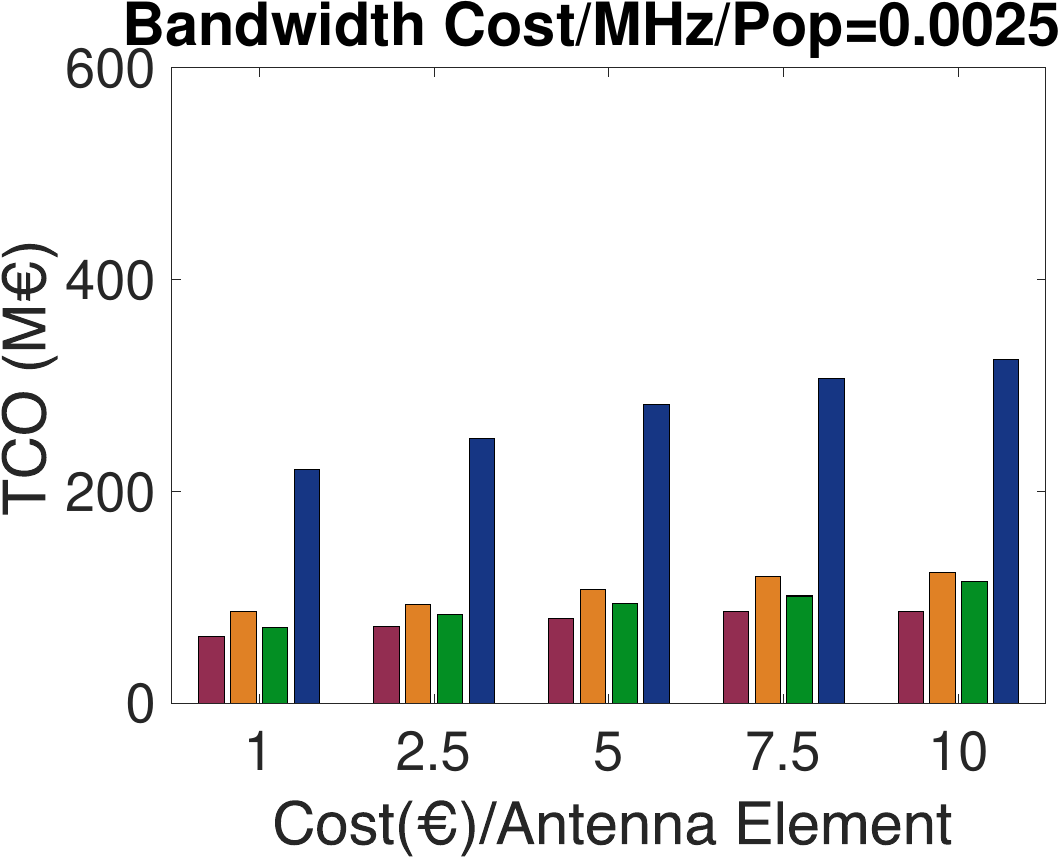}
        \end{subfigure}
        \begin{subfigure}[b]{0.19\textwidth}
                \centering
                \includegraphics[width=\textwidth]{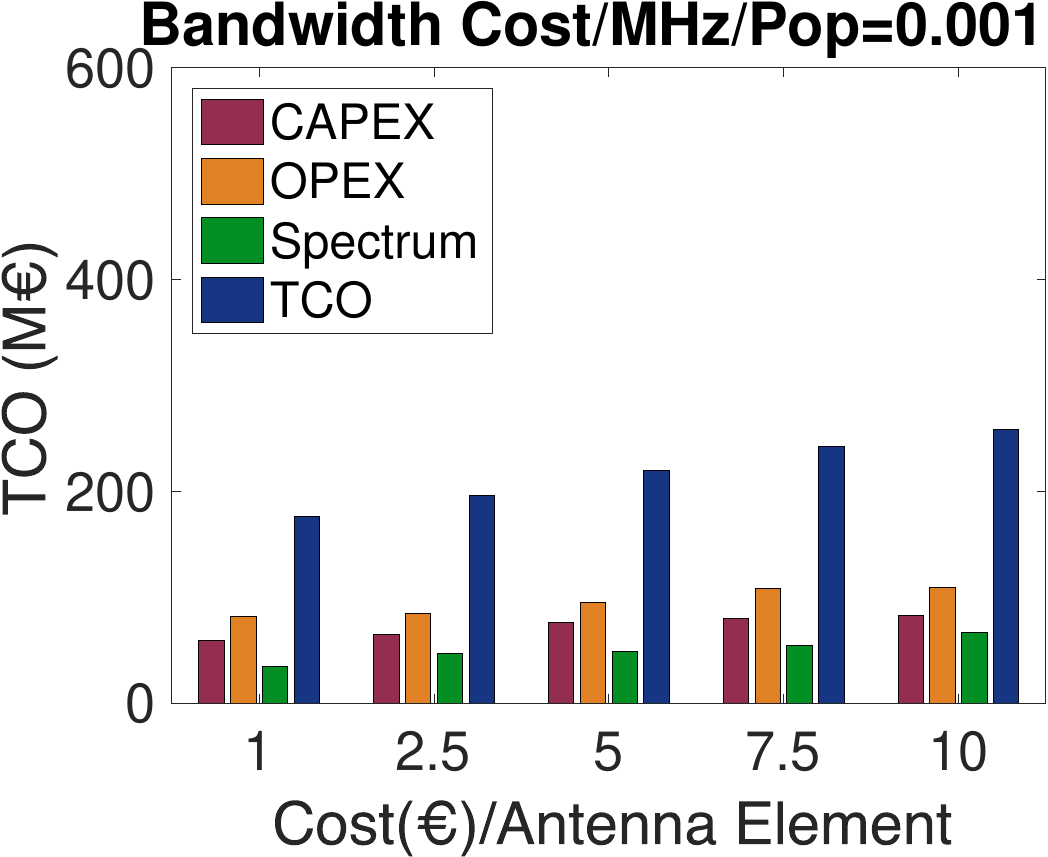}
        \end{subfigure}
        \caption{Distribution of TCO with different cost types for different antenna element and bandwidth costs to achieve $480$ Mbps.}
        \label{TCObar48}
\end{figure*}

\begin{figure*}[!ht]
        \centering
           \begin{subfigure}[b]{0.24\textwidth}
                \centering
                \includegraphics[width=\textwidth]{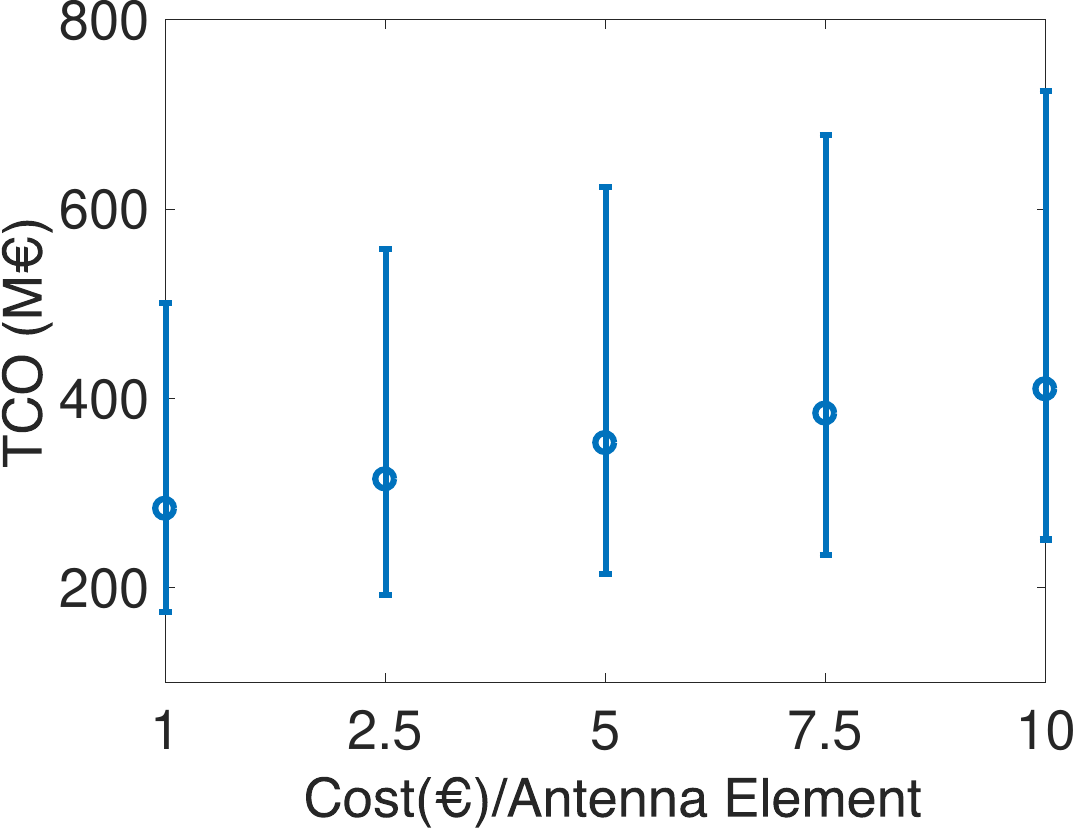}
                  \caption{$480$ Mbps.}
        \end{subfigure}
        \begin{subfigure}[b]{0.24\textwidth}
                \centering
                \includegraphics[width=\textwidth]{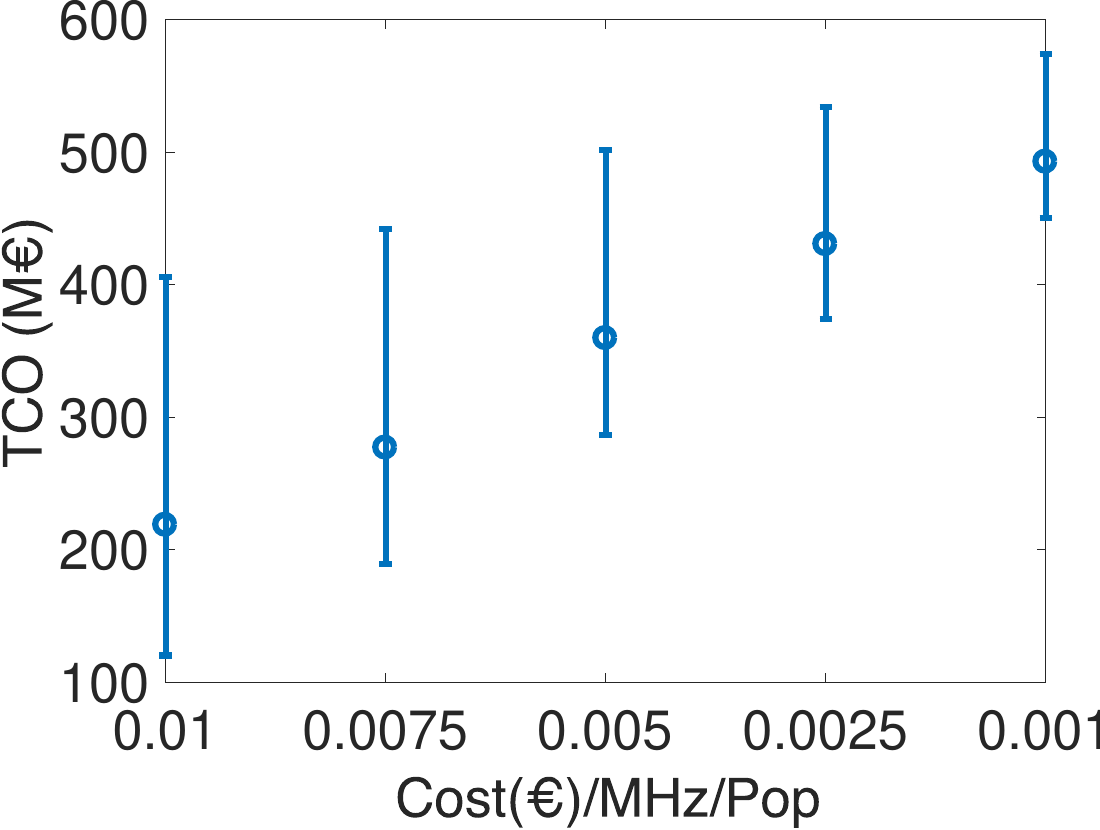}
                \caption{$480$ Mbps.}
        \end{subfigure}
         \begin{subfigure}[b]{0.24\textwidth}
                \centering
                \includegraphics[width=\textwidth]{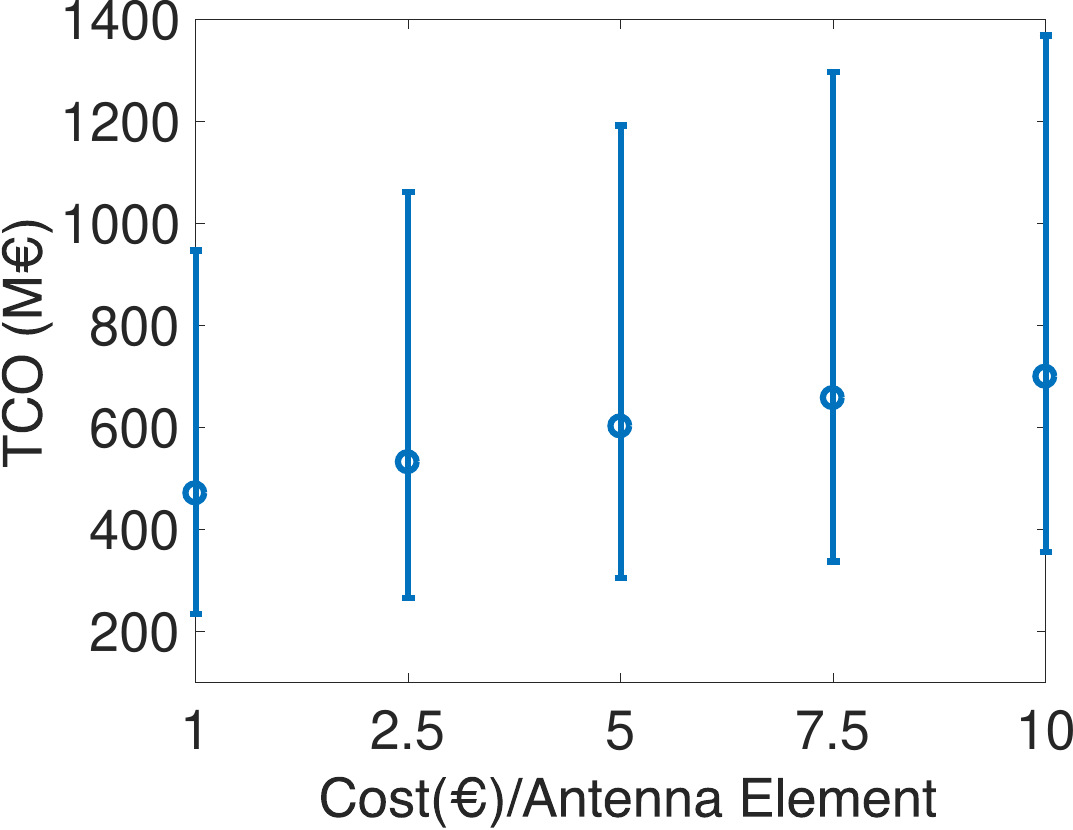}
                  \caption{$1.2$ Gbps.}
        \end{subfigure}
        \begin{subfigure}[b]{0.24\textwidth}
                \centering
                \includegraphics[width=\textwidth]{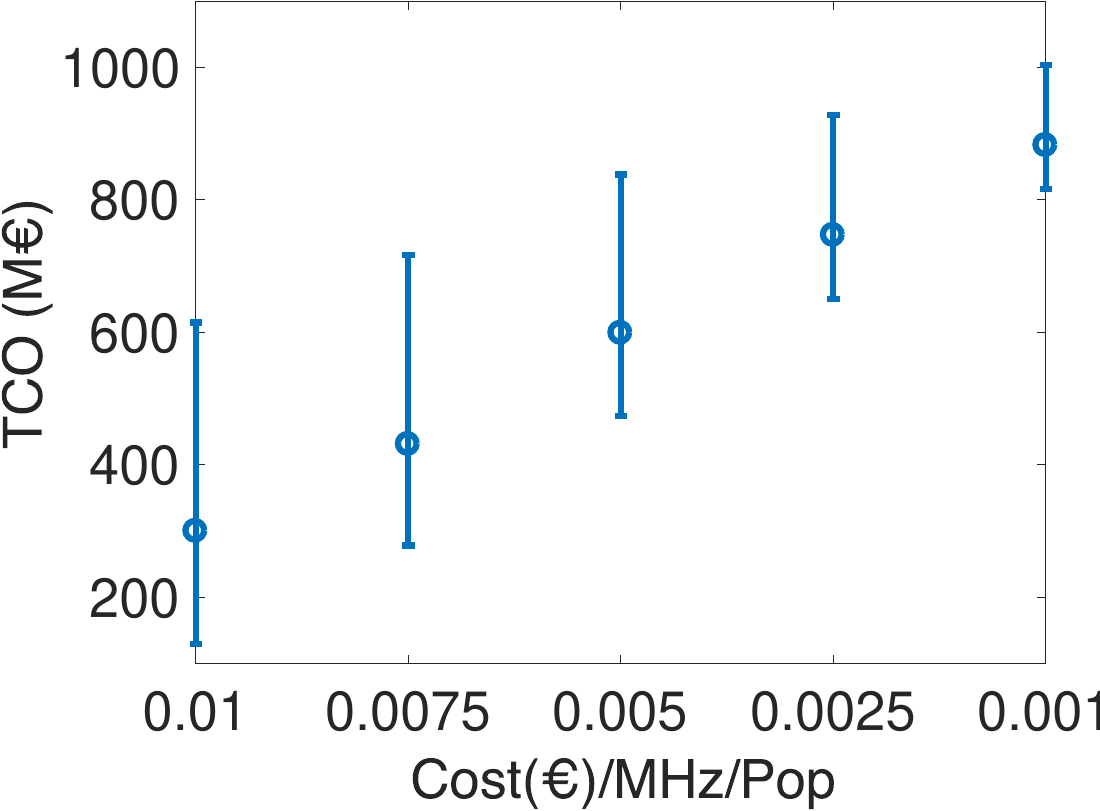}
                \caption{$1.2$ Gbps.}
        \end{subfigure}
        \caption{Deviation in TCO with $480$ Mbps and $1.2$ Gbps target data rate for (a, c) different antenna element costs with changing bandwidth cost, and (b, d) different bandwidth costs with changing antenna element cost.}
        \label{TCOerror12}
\end{figure*}

Fig. \ref{ac}(b,c) shows the number of total, and active aircraft ($K_{ac}$) in a cell. The simulation results are performed with the DFT-based elimination method as discussed in Section \ref{dftsubsec}. The analytical results are calculated with (\ref{findKAC}). As noticed, the number of aircraft in a cell is increasing with the cell range. In addition, {the increase in the number of aircraft is higher than the increase in the number of active aircraft as such the percentage of active aircraft decreases with the increasing range.} There are two reasons for this condition. The first is increasing number of aircraft in a cell, so that the probability of two aircraft being in the same beam increases. The second reason is that beam area becomes larger with increasing range. As noticed, the analytical results are consistent with the simulation results. 

Fig. \ref{rate} includes the results for mean aircraft data rate ($R^{est}_{Total}/K$). The simulation results are determined based on (\ref{actrtotal}) by calculating the beamforming matrices. (\ref{cellth}), (\ref{rhoest}) and (\ref{findKAC}) are utilized to calculate the analytical mean data rate results. We also compare the results with the no-interference case, in which the data rate is calculated by neglecting all the intra-cell interference. The analytical model and simulation results show high consistency especially higher than $50$ km range for all simulation scenarios. In order to keep the number of GS in practical levels, we focused on cell ranges between $50-100$ km.

\subsection{Ground Station Deployment}

The trade-off between the number of GSs and array size is presented in Fig. \ref{rate}(c) for $P_T=58$ dBm. For this result, we assume that the GS has four times more antenna elements than the aircraft side $N_T=4N_R$ because aircraft antenna size may have strict limitations due to aero-dynamical reasons. As noticed, availability of spectrum has a huge impact on the required number of antenna elements and number of GSs. Higher bandwidth levels require less number of antenna arrays and GSs. For example, for $900$ GSs deployed in Europe ($60$ km cell range), $50$ MHz requires $1296$ number of elements on the transmitter side, while $400$ and $256$ antenna elements are enough for $75$ MHz and $100$ MHz respectively. Most importantly, there is a saturation point such that the increase in the number of antenna elements do not provide significant decrease in the number of GSs. {Since traffic density is assumed as constant ($30$ aircraft/$18000$ km$^2$), adding more resources (GSs or antenna elements) do not provide additional gain after a certain point.} 

\subsection{TCO Optimization and Cost Sensitivity Analysis}

TCO optimization problem is given in Section \ref{susecprob}. To solve this problem in MATLAB, we utilize the fmincon and globalsearch functions together because the fmincon function converges to local minimum points. To tackle this problem, the globalsearch function utilize a dataset of points for the parameters. This way, the optimum points for the TCO problem can be calculated. We utilize the following loose upper and lower bounds for the variables $20$ km$\leq r_{max} \leq150$ km, $5$ $\leq N_T \leq 2L_T/\lambda$, $5$ $\leq N_R \leq 2L_R/\lambda$, $10$ dBm$\leq P_T
\leq 60$ dBm, $20$ MHz$\leq B\leq 200$ MHz.

Fig. \ref{TCObar12} and Fig. \ref{TCObar48} show the distribution of the TCO for different antenna element and bandwidth costs to achieve $1.2$ Gbps and $480$ Mbps, respectively. As noticed from the figures, the cost of spectrum does not only affect TCO, it also changes the result of the optimization problem as less number of GSs and antenna elements are required and the accumulated impact is visible in TCO reduction from sub-figures in Fig. \ref{TCObar12} and Fig. \ref{TCObar48}. The cost of antenna elements does not have a big effect on the overall TCO if the spectrum cost is low as a result of smaller number of antenna elements and GSs required for the deployment. It has larger impact when the spectrum is expensive. Therefore, the bandwidth cost is the most critical parameter to determine the cost of the network. 10-year OPEX of the network is also significant contributor for the TCO, and significant part of the OPEX is the maintenance and tower lease costs. Since the maintenance cost is calculated as 10\% of the CAPEX, it is also increasing with the antenna element cost. Thus, the OPEX costs increase when the number of GSs are high due to expensive bandwidth cost. In addition, Fig. \ref{TCOerror12} show the range of TCO with changing bandwidth costs and antenna element costs for achieving $1.2$ Gbps and $480$ Mbps, respectively. As noticed, the range is wider when the bandwidth cost is varying because the effect of the bandwidth cost is more dominant.

\begin{table*}[]
\centering
\caption{TCO Optimization Results.}
\label{TCOTable}
\scalebox{0.8}{
\begin{tabular}{|l|l|l|l|l|l|}
\hline
\multicolumn{6}{|c|}{$480$ Mbps}                                                                                                                                                                                             \\ \hline
\begin{tabular}[c]{@{}l@{}}Antenna element cost/\\ Bandwidth cost\end{tabular} & 1 $\euro{}$                & 2.5 $\euro{}$              & 5 $\euro{}$                & 7.5 $\euro{}$              & 10 $\euro{}$             \\ \hline
$0.01$ $\euro{}$/MHz/Pop     &$102.7$ km, ($3600$,$324$) &$107.1$ km, ($3600$,$169$) &$103.5$ km, ($2025$,$121$) &$97.27$ km, ($1681$,$81$) &$91.56$ km, ($1156$,$81$) \\ \hline
$0.0075$ $\euro{}$/MHz/Pop   &   $111.7$ km, ($3600$,$324$) &$114.4$ km, ($3600$,$169$) &$104$ km, ($1936$,$100$) &$98.72$ km, ($1444$,$81$) &$88.52$ km, ($961$,$64$)\\ \hline

$0.005$ $\euro{}$/MHz/Pop    & $116.3$ km, ($3600$,$225$) &$123.8$ km, ($3600$,$121$) &$118.3$ km, ($2209$,$81$) &$103.5$ km, ($1369$,$64$) &$109.7$ km, ($1156$,$64$)\\ \hline
$0.0025$ $\euro{}$/MHz/Pop       &$135.1$ km, ($3600$,$144$) &$140.4$ km, ($3364$,$100$) &$125$ km, ($1849$,$64$) &$114.9$ km, ($1296$,$49$) &$109.7$ km, ($900$,$36$)\\ \hline

$0.001$ $\euro{}$/MHz/Pop      &$138.7$ km, ($2704$,$100$) &$144.4$ km, ($2116$,$64$) &$144.4$ km, ($2401$,$49$) &$123.2$ km, ($1296$,$36$) &$125.3$ km, ($961$,$36$)\\ \hline
\multicolumn{6}{|c|}{$1.2$ Gbps}                                                                                                                                                                                             \\ \hline
\begin{tabular}[c]{@{}l@{}}Antenna element cost/\\ Bandwidth cost\end{tabular} & 1 $\euro{}$                & 2.5 $\euro{}$              & 5 $\euro{}$                & 7.5 $\euro{}$              & 10 $\euro{}$             \\ \hline
$0.01$ $\euro{}$/MHz/Pop 
 &$89.06$ km, ($3600$,$625$) &$88.31$ km, ($3481$,$324$) &$84.44$ km, ($2601$,$196$) &$79.84$ km, ($1764$,$169$) &$77.86$ km, ($1444$,$144$)\\ \hline
$0.0075$ $\euro{}$/MHz/Pop                                                   
 &$89.06$ km, ($3600$,$576$) &$86.6$ km, ($3600$,$289$) &$79.05$ km, ($1764$,$196$) &$75$ km, ($1225$,$121$) &$75.17$ km, ($1296$,$121$)\\ \hline
$0.005$ $\euro{}$/MHz/Pop                              
 &$95.61$ km, ($3600$,$441$) &$103$ km, ($3600$,$225$) &$89.39$ km, ($2025$,$144$) &$85.48$ km, ($1521$,$121$) &$79.05$ km, ($961$,$100$)\\ \hline
$0.0025$ $\euro{}$/MHz/Pop                            
 &$112.4$ km, ($3600$,$324$) &$108.7$ km, ($3600$,$169$) &$99.17$ km, ($1936$,$100$) &$101.3$ km, ($1764$,$100$) &$97.84$ km, ($1156$,$81$)\\ \hline
$0.001$ $\euro{}$/MHz/Pop            
 &$132.9$ km, ($3600$,$196$) &$131.8$ km, ($3600$,$121$) &$115.4$ km, ($1936$,$81$) &$104.2$ km, ($1225$,$49$) &$104.2$ km, ($1089$,$49$)\\ \hline
\end{tabular}}
\end{table*}


Table \ref{TCOTable} presents the network parameters to achieve $480$ Mbps and $1.2$ Gbps data rates. Since the contribution of the power consumption is significantly lower than other factors, all the cases have the maximum power budget of $60$ dBm. For $480$ Mbps, the cell ranges are changing between $90-145$ km, and the number of antenna elements in the GS side is between $900-3600$ whereas AS side has much simpler antenna array between $36-324$. In the case of $1.2$ Gbps, there is a need for more GSs with the cell ranges between $75-132$ km. The number of antenna elements for GS and AS sides is changing between $961-3600$ and $49-625$, respectively. As noticed from these results, the optimal points are achieved by the GS with high number of antenna elements to lower the operational costs: maintenance and lease. Since the number of ASs are significantly higher than the number of GS, AS side has lower antenna array sizes to minimize TCO.  


\section{Conclusion}
\label{con}

In this paper, we address the GS deployment problem and TCO optimization to calculate the main DA2GC design parameters: the number of GSs, the number of antenna elements, bandwidth and transmit power. To this end, we propose a multi-user beamforming algorithm for dual-polarized antenna arrays, and develop an analytical expression for the DA2GC cell throughput. In addition, the GS antenna structure is proposed to lower the beamsteering-loss, and according to our analysis, $8$ faceted structure provides the optimal solution. TCO optimization is performed to reach $480$ Mbps and $1.2$ Gbps mean capacity for different bandwidth and antenna element costs. {The interplay between different network resources and their relative cost is demonstrated via sensitivity analysis.} The investigation has revealed that the bandwidth cost is the dominant factor. On the other hand, the contribution of power consumption is significantly lower than the contribution of other factors, and maximum allowed transmission power can be utilized by GSs. The proposed framework in this paper can be further extended to include actual flight routes. {In addition,  the power consumption in the network can be further improved by assuming sleep mode when there is no aircraft in the vicinity of a GS.}




\begin{thebibliography}{1}

\bibitem{vtc}
E. Dinc, M. Vondra, C. Cavdar, ``Seamless Gate-to-Gate Connectivity Concept: Onboard LTE, Wi-Fi and LAA," in \emph{Proc. IEEE VTC 2017-Fall}, Toronto, Canada, September 2017 [online].

\bibitem{dinc2017commag}
E. Dinc et al. ``In-flight Broadband Connectivity: Architectures and Business Models for High Capacity Air-to-Ground Communications", \emph{IEEE Communications Magazine}, vol. 55, no. 9, pp. 142-149, 2017.

 \bibitem{atag}
Air Transport Action Group, ``Aviation: Benefits Beyond Borders", Jul. 2016.


 
 \bibitem{gogo2ku}
B. R. Elbert, ``Aeronautical Broadband for Commercial Aviation: Evaluating the 2Ku Solution, Application Technology Strategy," L.L.C., Oct. 2014.

\bibitem{vondra17c}
M. Vondra, E. Dinc, M. Prytz, M. Frodigh, D. Schupke, M. Nilson, S. Hofmann, C. Cavdar, ``Performance Study on Seamless DA2GC for Aircraft Passengers toward 5G", \emph{IEEE Communications Magazine}, vol. 55, no. 11, pp. 194-201, Nov. 2017



\bibitem{atg4}
Gogo Inc., ATG-4 [online]. Available: http://concourse.gogoair.com/technology/gogo-atg-4-work

\bibitem{inmarsat}
Inmarsat, ``The European Aviation Network," [online]. Available: https://www.telekom.com/static/-/288318/2/150921-product-sheet-si

\bibitem{ecc}
CEPT ECC Report 214, 6 Jun. 2014. ``Broadband Direct-Air-to-Ground Communications (DA2GC)".





 \bibitem{ngmn}
NGMN Alliance, ``5G White Paper," Feb. 2015.


\bibitem{mozaffari17}
M. Mozaffari et al.,, ``Wireless Communication Using Unmanned Aerial Vehicles (UAVs): Optimal Transport Theory for Hover Time Optimization," \emph{IEEE Transactions on Wireless Communications, vol. 16, no. 12, pp. 8052-8066, Dec. 2017.}

\bibitem{khuwaja18} 
A. A. Khuwaja, Y. Chen, N. Zhao, M. Alouini and P. Dobbins, ``A Survey of Channel Modeling for UAV Communications," \emph{IEEE Communications Surveys \& Tutorials}, vol. 20, no. 4, pp. 2804-2821, Fourthquarter 2018.

\bibitem{khawaja18}
W. Khawaja, I. Guvenc, D. Matolak, U. C. Fiebig, and N. Schneckenberger, ``A survey of air-to-ground propagation channel modeling for unmanned aerial vehicles," \emph{arXiv preprint arXiv:1801.01656}, 2018.	

\bibitem{qureshi18}
H. N. Qureshi and A. Imran, "Towards Designing Systems with Large Number of Antennas for Range Extension in Ground-to-Air Communications," in \emph{Proc. of IEEE PIMRC}, Bologna, 2018, pp. 1-5.

\bibitem{tadayon16}
N. Tadayon, G. Kaddoum and R. Noumeir, ``Inflight Broadband Connectivity Using Cellular Networks," \emph{IEEE Access}, vol. 4, pp. 1595-1606, 2016.


\bibitem{tang18}
H. Tang et al., "Massive MIMO Antenna Array Deployment for Airport in Air-to-Ground Communications," 2018 Cross Strait Quad-Regional Radio Science and Wireless Technology Conference (CSQRWC), Xuzhou, 2018, pp. 1-2.


\bibitem{helfrick07}
A. Helfrick, Principles of Avionics. Leesburg, VA, USA: Avionics Communications Inc., 2007.
 
  \bibitem{sesar}
``State of progress with the project to implement the new generation European Air Traffic Management System (SESAR)," European Union, Brussels, Belgium, 2007.

\bibitem{plass12}
S. Plass, ``Seamless networking for aeronautical communications: One major aspect of the SANDRA concept," in \emph{IEEE Aerospace and Electronic Systems Magazine}, vol. 27, no. 9, pp. 21-27, September 2012.
 

\bibitem{schnell04}
M. Schnell, E. Haas, M. Sajatovic, C. Rijacek, and B. Haindl, ``B-VHF: An overlay system concept for future ATC communications in the VHF band," in \emph{Proc. DASC}, 2004,  pp. 1-9.
 
 \bibitem{ehammer11}
M. Ehammer, E. Pschernig, and T. Graupl, "AeroMACS: An Airport Communications System," in \emph{IEEE/AIAA E Digital Avionics Systems Conference (DASC)}, 2011.

\bibitem{fernandez15}
D. Fernandez, M. l. Admella, L. Albiol and J. M. Cebrian, ``Satellite Communications Data Link Solution for Long Term Air Traffic Management," \emph{Fourth SESAR Innovation Days}, November 2014.    

\bibitem{dinc17}
E. Dinc, M. Vondra, C. Cavdar, ``Multi-user Beamforming and Ground Station Deployment Problem for 5G Direct Air-to-Ground Communications," to appear in {\em Proc. of IEEE GLOBECOM}, 2017 [online]. Available: https://goo.gl/PUAjGe

\bibitem{pi16}
Z. Pi, J. Choi, R. Heath, ``Millimeter-wave gigabit broadband evolution toward 5G: fixed access and backhaul," \emph{IEEE Communications Magazine}, vol. 54, no. 4, pp. 138-144, April 2016.


                                	
\bibitem{tan15}
W. Tan, S. Jin, J. Wang, M. Matthaiou, ``Achievable sum-rate of multiuser massive MIMO downlink in ricean fading channels," in \emph{Proc. of ICC}, London, 2015, pp. 1453-1458.
 



 
 






\bibitem{dualpol}
K. P. Liolis, J. Gomez-Vilardebo, E. Casini, A. I. Perez-Neira, ``Statistical Modeling of Dual-Polarized MIMO Land Mobile Satellite Channels," \emph{IEEE Trans. on Comm.}, vol. 58, no. 11, pp. 3077-3083, Nov. 2010.

\bibitem{chang15}
D. Chang, J. Lee, ``Satellite communications service provision via incompatible polarization diversity and orthogonal beams," in \emph{Proc. of WCNC}, New Orleans, LA, 2015, pp. 2256-2261.

\bibitem{ADSB2020}
Department of Transportation Federal Aviation Administration, 14 CFR Part 91, ``Automatic Dependent Surveillance Broadcast (ADS-B) Out Performance Requirements To Support ATC Service", May 28, 2010

\bibitem{richards2014} 
R. William, K. O'Brien, C. Miller, ``New Air Traffic Surveillance Technology" (PDF). Boeing aero quarterly, 7 April 2014 [online]. Available: https://goo.gl/nv976s

\bibitem{kim16}
Y. Kim et al., "A secure location verification method for ADS-B," in \emph{Proc. of IEEE/AIAA DASC}, CA, 2016, pp. 1-10.


\bibitem{you13}
H. You, Z. Hongwei and T. Xiaoming, ``Joint systematic error estimation algorithm for radar and automatic dependent surveillance broadcasting," \emph{IET Radar, Sonar \& Navigation}, vol. 7, no. 4, pp. 361-370, April 2013.


\bibitem{nijsure16_1}
Y. A. Nijsure et al., ``Adaptive Air-to-Ground Secure Communication System Based on ADS-B and Wide-Area Multilateration," \emph{IEEE Transactions on Vehicular Technology}, vol. 65, no. 5, pp. 3150-3165, May 2016.

\bibitem{nijsure16_2}
Y. Nijsure, M. F. A. Ahmed, G. Kaddoum, G. Gagnon and F. Gagnon, ``WSN-UAV Monitoring System with Collaborative Beamforming and ADS-B Based Multilateration," in \emph{VTC Spring-2016}, Nanjing, 2016, pp. 1-5.



\bibitem{emil}
E. Bjornson, et al., ``Massive MIMO: ten myths and one critical question,"  \emph{IEEE Commun. Mag.}, vol. 54, no. 2, Feb. 2016.

\bibitem{array} 
E. Brookner, ``Phased Array Antenna Handbook," Artech House, 2nd ed., sec. 3, 2005.





\bibitem{sp7}
C. Doan et al., ``Design considerations for 60 GHz CMOS radios," \emph{IEEE Commun. Mag.}, vol. 42, no. 12, 2004.

\bibitem{sp8}
Z. Pi et al. ``An introduction to millimeter-wave mobile broadband systems," \emph{IEEE Commun. Mag.}, vol. 49, no. 6, 2011.

\bibitem{mult}
A. Alkhateeb, et al., ``Limited Feedback Hybrid Precoding for Multi-User Millimeter Wave Systems," \emph{IEEE Trans. on Wireless}, vol. 14, no. 11, pp. 6481-6494, Nov. 2015.

\bibitem{xia}
P. Xia, et al., ``A practical SDMA protocol for 60 GHz millimeter wave communications," in \emph{Proc. 2008 Asilomar Conf.}, pp. 2019-2023.

\bibitem{xia2}
P. Xia, et al., ``Multi-stage iterative antenna training for millimeter wave communications," in \emph{Proc. GLOBECOM}, LA, USA, 2008, pp. 1-6.

\bibitem{spencer04}
Q. H. Spencer, A. L. Swindlehurst, M. Haardt, ``Zero-forcing methods for downlink spatial multiplexing in multiuser MIMO channels," \emph{IEEE Trans. on Signal Process.}, vol. 52, no. 2, pp. 461-471, Feb. 2004.

\bibitem{kim09}
J. Kim, D. J. Park, ``Constrained Codebook Design for a MISO Beamforming System in a Rician Channel," in \emph{Proc. of VTC-Spring}, 2009, pp. 1-5.


\bibitem{qureshi19}
H. N. Qureshi and A. Imran, ``On the Tradeoffs between Coverage Radius, Altitude and Beamwidth for Practical UAV Deployments," \emph{IEEE Transactions on Aerospace and Electronic Systems}, 2019


\bibitem{ibmanten}
X. Gu, et. al, ``A Multilayer Organic Package with 64 Dual-Polarized Antennas for 28GHz 5G Communication," in \emph{Proc. of IMS}, 2017. 

\bibitem{jan14}
A. A. Widaa Ahmed, K. Chatzimichail, J. Markendahl and C. Cavdar, ``Techno-economics of Green Mobile Networks Considering Backhauling," in \emph{Proc. of European Wireless 2014}, Barcelona, Spain, 2014, pp. 1-6.

\bibitem{jan142}
A. A. W. Ahmed, J. Markendahl and C. Cavdar, ``Interplay between cost, capacity and power consumption in heterogeneous mobile networks," in \emph{Proc. of ICT 2014}, Lisbon, 2014, pp. 98-102.

\bibitem{tower}
``New site typologies and business models for 5G," in \emph{Journal of the telecom tower industry in EMEA, CALA and Asia}, iss. 18, Jan. 2017 [online]. Available: https://www.towerxchange.com/wp-content/uploads/2017/02/TowerXchange-Issue\_18.pdf

\bibitem{jan3}
J. Markendahl and O. Makitalo, ``A comparative study of deployment options, capacity and cost structure for macrocellular and femtocell networks," in \emph{Proc. of PIMRC 2010,} Istanbul, 2010, pp. 145-150.

\bibitem{hybridRF}
A. Garcia-Rodriguez, V. Venkateswaran, P. Rulikowski and C. Masouros, ``Hybrid Analog-Digital Precoding Revisited Under Realistic RF Modeling," in \emph{IEEE Wireless Communications Letters}, vol. 5, no. 5, pp. 528-531, Oct. 2016.

\bibitem{emil15}
E. Björnson, L. Sanguinetti, J. Hoydis and M. Debbah, "Optimal Design of Energy-Efficient Multi-User MIMO Systems: Is Massive MIMO the Answer?," \emph{IEEE Transactions on Wireless Communications}, vol. 14, no. 6, pp. 3059-3075, June 2015.

\bibitem{paeeff}
J. Zhao, M. Bassi, A. Mazzanti and F. Svelto, ``A 15 GHz-bandwidth 20dBm PSAT power amplifier with 22\% PAE in 65nm CMOS," in \emph{Proc. of CICC 2015,}, San Jose, CA, 2015, pp. 1-4.

\bibitem{kwhprice}
http://ec.europa.eu/eurostat/statistics-explained/index.php/File:Half-yearly\_electricity\_prices\_(EUR).png

\bibitem{janspec}
B. G. Molleryd and J. Markendahl, ``Valuation of spectrum for mobile broadband services - Engineering value versus willingness to pay," in \emph{22nd European Regional ITS Conference,} Budapest, 18-21 Sep., 2011.

\bibitem{kavanagh18}
S. Kavanagh, ``5G UK auction," 5G.co.uk, 2018 [online]. Available: https://5g.co.uk/guides/5g-uk-auction/




\end{thebibliography}
\end{document}